\documentclass[12pt]{article}
\pdfoutput=1

\usepackage[all]{xy}

\usepackage{xcolor}
\usepackage{dsfont}
\usepackage{tabularx}

\usepackage{hyperref}
\usepackage{subcaption}
\usepackage[utf8]{inputenc}
\usepackage{ytableau}
\usepackage[vcentermath]{youngtab}

\usepackage{setspace}

\usepackage{amsmath, amssymb, amsthm, float, graphicx}
\numberwithin{equation}{section}
\usepackage{booktabs}

\hypersetup{colorlinks=false, linkcolor=blue, citecolor=red}

\def\beq{\begin{eqnarray}}\def\eeq{\end{eqnarray}}
\def\be{\begin{equation}}\def\ee{\end{equation}}

\def\zb{\bar{z}}

\def\a{\alpha}
\def\e{\epsilon}
\def\k{\kappa}
\def\b{\beta}
\def\d{\delta}

\def\D{\Delta}

\newcommand{\mf}[1]{\mathfrak #1}

\begin{document}
	
	\title{\bf Crossing antisymmetric Polyakov blocks + Dispersion relation}
		\date{}
	\maketitle
	

	\begin{center}
		{\bf 
			Apratim Kaviraj
		} 
		\\
		\vspace{.2in} 
		{\it DESY Hamburg, Theory Group, \\ Notkestra\ss e 85, D-22607 Hamburg, Germany}\\
	\end{center}

	\begin{center}
	{ \ \ \texttt{apratim.kaviraj@desy.de} \ \  
	}
	\\
\end{center}
	
	\vspace{.2in}

	\begin{abstract}
		Many CFT problems, e.g. ones with global symmetries, have correlation functions with a crossing antisymmetric sector. 
		We show that such a  crossing antisymmetric function can be expanded in terms of manifestly crossing antisymmetric objects, which we call the `$+$ type Polyakov blocks'. These blocks are built from AdS$_{d+1}$ Witten diagrams. In 1d they encode the `$+$ type' analytic functionals which act on crossing antisymmetric functions. In general $d$ we establish this Witten diagram basis from a crossing antisymmetric dispersion relation  in Mellin space.  Analogous to the crossing symmetric case, the dispersion relation imposes a set of independent  `locality constraints' in addition to the usual CFT sum rules given by the `Polyakov conditions'. We use the Polyakov blocks to simplify more general analytic functionals in $d>1$ and global symmetry functionals.

	\end{abstract}
	\vspace{.2in}
	\vspace{.3in}
	\clearpage
	
	\tableofcontents

	\section{Introduction}\label{sec:intro}

	One of the most important tools available to theoretical physicists in CFT problems is crossing symmetry or the Conformal Bootstrap. For numerical analysis one typically approaches the 4-point crossing problem by introducing a set of functionals, which are simply a set of derivatives w.r.t. the cross ratios.  From the recent works of \cite{Mazac:2016qev, Mazac:2018mdx, Mazac:2018ycv} (see also \cite{El-Showk:2016mxr} and futher generalizations \cite{Kaviraj:2018tfd,Mazac:2018biw,Mazac:2019shk,Caron-Huot:2020adz,Giombi:2020xah}) we have learnt that there exists a more efficient set of functionals, the  analytic functionals.  These are built to optimize OPE coefficients, and are `dual' to GFF or Generalized Free Fields (i.e. a  basis expansion in terms of double trace conformal blocks). 
	
	The analytic functionals are perhaps most well understood in 1d, where they are related to another formulation of crossing symmetry: the Polyakov bootstrap \cite{Polyakov}. The latter is an idea that a crossing symmetric correlator can be expanded in terms of a manifestly crossing symmetric sum of Witten diagrams (a Polyakov block) which must be equal to the usual conformal block expansion \cite{Sen:2015doa,Gopakumar:2016cpb,Gopakumar:2016wkt}. The validity of this has been rigorously proven by the 1d analytic functionals. In 1d, Polyakov blocks also give a simple way to determine action of the functionals, which are otherwise hard to compute. This connection has been recently used in \cite{Ghosh:2021ruh} for general global symmetries to obtain a number of interesting numerical results.

	Recently in \cite{Sinha:2020win,Gopakumar:2021dvg} a crossing symmetric dispersion relation has been proposed which establishes the Polyakov Bootstrap in general $d$. The main idea here, which is based on \cite{Auberson:1972prg}, is to work in Mellin space and map the Mellin variables  to new variables that are manifestly crossing symmetric but with nonlocal singularities. A dispersion relation is then obtained in these variables, which reduces to a crossing symmetric sum of exchange and contact Witten diagrams when one sets the nonlocal singularities to zero (locality constraints). The dispersion relation is also used in the QFT context (replacing Mellin variables with Mandelstam variables) where these locality constraints have been used to derive interesting bounds on Effective Field Theories (EFTs).
	
	
	Let us now address a situation where crossing symmetry of a 4-point function involves more than just crossing symmetric functions. E.g.
	consider a CFT with global symmetry with a correlator $\mathcal{G}_{ijkl}(z,\zb)$ of charged fields $\phi_i$ transforming in some irrep $\mf r$. If $\mf r \otimes \mf r$ contains the irreps labelled $\mf a$ then the correlator decomposes as
	\be\label{eq:Gijkl}
	\mathcal{G}_{ijkl}(z,\zb)= \sum_{\mf a}T^{\mf a}_{ij,kl}\mathcal{G}^{\mf a}(z,\zb)
	\ee
	where $T^{\mf a}$ denotes the associated tensor structure of $\mf a$. We use the notations of  usual CFT kinematics, reviewed in section \ref{sec:1dfunc}.
	The crossing equation for $	\mathcal{G}^{\mf a}$ can be written as:
	\be\label{crossing}
	\mathcal{G}^{\mf a}(z,\zb)=C^{\mf a \mf b} \mathcal{G}^{\mf b}(1-z,1-\zb)\,.
	\ee
	Here $C^{\mf a\mf b}$ is a crossing matrix (see e.g. \cite{Ghosh:2021ruh}) that has eigenvalues $\pm 1$ corresponding to eigenvectors $e^{\mf a}_{\pm,s}$ where $s$ is a label.  Now, while the functions $\mathcal{G}_{-}(z,\zb)=e^{\mf a}_{+,s}\mathcal{G}^{\mf a}(z,\zb)$, for any $s$, obey usual crossing symmetry $\mathcal{G}_{-}(z,\zb)=\mathcal{G}_{-}(1-z,1-\zb)$, the combinations $\mathcal{G}_{+}(z,\zb)=e^{\mf a}_{-,s}\mathcal{G}^{\mf a}(z,\zb)$ are functions that have crossing antisymmetry:
	\be\label{eq:+typesol}
	\mathcal{G}_{+}(z,\zb)=-\mathcal{G}_{+}(1-z,1-\zb)\,.
	\ee
	An obvious question to ask is: does $\mathcal{G}_+$ also allow a Witten diagram like expansion?
 In 1d there exists a set of analytic functionals,  called $+$ type functionals, which can bootstrap such antisymmetric functions although it is not understood if they are related to a Polyakov block.\footnote{In \cite{Ghosh:2021ruh} this was bypassed by considering analytic functionals dual to GFF for a certain global symmetry which are related to crossing symmetric Polyakov blocks of the same symmetry \cite{Dey:2016mcs,Ferrero:2019luz}.} It would be nice to understand this more generally in $d\ge 1$. 
	
	In this paper we show that, analogous to the crossing symmetric case,  crossing antisymmetric functions can be expanded in terms of manifestly crossing antisymmetric objects as follows:
	\be
	\mathcal{G}_+(z,\zb)=\sum_{\D,\ell}a_{\D,\ell}\mathcal{P}_{+,\D,\ell}(z,\zb)\,, 
	\ee
	where $\mathcal{P}_{+,\D,\ell}$, the `$+$ type Polyakov block', is a linear combination of AdS$_{d+1}$ Witten diagrams.  In 1d these blocks involve a finite number of diagrams similar to its crossing symmetric counterpart. They are related to the $+$ type analytic functionals. We construct two sets of blocks corresponding to the bosonic and fermionic functional bases. 
	
	The idea of a  $+$ type Polyakov block is then extended to general dimensions, which requires an infinite number of exchange  Witten diagrams and crossing antisymmetric contact diagrams. For this we show that a crossing antisymmetric function can be obtained from a dispersion relation in Mellin space. Analogous to the crossing symmetric case, it  is characterized by an infinite number of nonlocal singularities that lead to a new set of locality constraints. The dispersion relation can be reformulated as the Witten diagram expansion once the locality conditions are satisfied.  Finally we give a set of crossing antisymmetric sum rules.  
	
	We have numerically demonstrated how locality conditions work, and how a known crossing antisymmetric solution satisfies the new sum rules.
	We show how the  $+$ type Polyakov blocks simplify computation of `product functionals' which exist in even dimensions and are built by combining 1d functionals \cite{Paulos:2019gtx}. 
	We also show one can obtain a set of `simple  functionals' for global symmetry problems  in general dimensions  from the Polyakov blocks extending the arguments of \cite{Ghosh:2021ruh}.
	
	The paper is organized as follows: in section \ref{sec:1dfunc} we review the $\pm$ type 1d functionals and introduce the $+$ type bosonic and fermionic Polyakov blocks. In section \ref{sec:gendPolyblock} we generalize the $+$ type Polyakov blocks to arbitrary dimensions and introduce a crossing antisymmetric dispersion relation. In section \ref{sec:applications} we show how the previous findings simplify various $d\ge 1$ functionals. We conclude in section \ref{sec: conclusion}. There are four  appendices elaborating on the many technical details and numerical checks. 
	
	\textbf{Notations:} Throughout this paper we will use the subscripts $-$/$+$ to indicate  crossing symmetric/antisymmetric equations or quantities. This, rather counter-intuitive, notation is for consistency with the analytic functional literature.

	\section{1d crossing antisymmetric functionals}\label{sec:1dfunc}
	
	\subsection{Basic kinematics}
	
	Let us define some basic kinematics and notations. Consider a 4-point function $\mathcal{G}(x_1,x_2,x_3,x_4)$ of scalars (say $\phi(x_i)$) with identical dimension $\Delta_\phi$ in a $d$-dimensional CFT which is given by 
	\be
     \mathcal{G}(x_1,x_2,x_3,x_4)=\frac{\mathcal{G}(z,\zb)}{x_{13}^{2\D_\phi}x_{24}^{2\D_\phi}}\,, \ \ z \zb=\frac{x_{12}^2x_{34}^2}{x_{13}^2x_{24}^2}\,, \ (1-z)(1-\zb)=\frac{x_{14}^2x_{23}^2}{x_{13}^2x_{24}^2}\,.
	\ee
	We will sometimes use the notations $u=z \zb$ and $v=(1-z)(1-\zb)$. The quantity $\mathcal{G}(z,\zb)$ can be decomposed into conformal blocks as follows:
	\be\label{confblockdeco}
	\mathcal{G}(z,\zb)=\sum_{\D,\ell}a_{\D,\ell} G_{\Delta,\ell}^{(d)}(z,\zb|\D_\phi)\,.
	\ee
	Here $G_{\Delta,\ell}^{(d)}(z,\zb|\D_\phi)$ is the $d$-dimensional conformal block (defined with a factor of $(z\zb)^{-\D_\phi}$) and $a_{\D,\ell}$ denotes square of OPE coefficients. Typically the function $\mathcal{G}(z,\zb)=\mathcal{G}_-(z,\zb)$ is crossing symmetric i.e. $\mathcal{G}_-(z,\zb)=\mathcal{G}_-(1-z,1-\zb)$ (symmetry under $x_1\leftrightarrow x_3$ exchange) and we may write  
	\be
	\sum_{\D,\ell}a_{\D,\ell} F_{-,\D,\ell}(z,\zb|\D_\phi)=0
	\ee
	where we defined the crossing symmetric bootstrap vector
	\be
	F_{-,\D,\ell}(z,\zb|\D_\phi)=G_{\D,\ell}(z,\zb|\D_\phi)-G_{\D,\ell}(1-z,1-\zb|\D_\phi)\,.
	\ee
	As shown in \eqref{eq:+typesol} it is often necessary to consider  $\mathcal{G}(z,\zb)=\mathcal{G}_+(z,\zb)$ which is antisymmetric under $x_1\leftrightarrow x_3$ exchange i.e. $\mathcal{G}_+(z,\zb)=-\mathcal{G}_+(1-z,1-\zb)$. If it decomposes into conformal blocks like \eqref{confblockdeco} we have the crossing antisymmetry equation
	\be
	\sum_{\D,\ell}a_{\D,\ell} F_{+,\D,\ell}(z,\zb|\D_\phi)=0
	\ee
	where we have the crossing antisymmetric bootstrap vector
	\be
	F_{+,\D,\ell}(z,\zb|\D_\phi)=G_{\D,\ell}(z,\zb|\D_\phi)+G_{\D,\ell}(1-z,1-\zb|\D_\phi)\,.
	\ee
	We will sometimes loosely refer to the crossing antisymmetric equation as `anticrossing equation'. It becomes important in many interesting problems like bootstrapping CFTs with global symmetries or multiple correlators. 
	
	For the rest of this section we will focus on 1d CFTs. Here we have a single cross-ratio, so we set $z=\zb$. The (anti)crossing equation is written as
	\be
	\sum_{\D}a_{\D} F_{\pm,\D}(z|\D_\phi)=0
	\ee
	where we have 
	\be
	F_{\pm}(z|\D_\phi)=G_{\D}(z|\D_\phi)\pm G_{\D}(1-z|\D_\phi)
	\ee
	and $G_{\D}(z|\D_\phi)=z^{\D-2\D_\phi}{}_2F_1(\D,\D,2\D,z)$ is the $SL(2,\mathbb{R})$ conformal block\,.
	
	\subsection{1d analytic functionals and Polyakov Bootstrap}
	The usual approach to constrain CFT data from the (anti)crossing equation is by introducing a set of linear functionals as follows ($\omega(\D|\D_\phi)\equiv \omega[F_{\D}(z|\D_\phi)] $)
	\be
	\omega_\pm\Big[\sum_{\D}a_\D F_{\pm,\D}(z|\D_\phi)\Big]= \sum_{\D}a_\D\omega_\pm (\D|\D_\phi)=0
	\ee
	In the standard conformal bootstrap approach one chooses a basis of derivatives $\omega_{\pm}=\{\partial_z,\partial_z^2,\cdots \}$ at $z=\frac 12$\,. In \cite{Mazac:2018mdx,Mazac:2018ycv} a set of functionals that gives optimal bounds on OPE coefficients were proposed. These are the analytic functionals which correspond to a basis expansion of bootstrap vectors in terms of double trace operator blocks as follows:
	\be\label{eq:basis}
	F_{\pm,\D}(z|\D_\phi)=\sum_{n}\Big[\a^{B,F}_{\pm,n}(\D)\, F_{\pm,\D_{n}^{B,F}}(z|\D_\phi)+\b^{B,F}_{\pm,n}\, \partial_\D F_{\pm,\D_{n}^{B,F}}(z|\D_\phi)\Big]\,.
	\ee
	Here we have shown two set of bases:  Bosonic ($B$) and Fermionic ($F$)  which correspond to $\D_n^B=2\D_\phi+2n$ and  $\D_n^F=2\D_\phi+2n+1$. Each analytic functional $\omega_\pm=\a_\pm,\b_\pm$ is dual to a basis element above. It is defined as follows:
	\be\label{funcdef}
	\omega_\pm(\D|\D_\phi)=\int_{\frac 12}^{\frac 12+i\infty}dz\, f_{\pm}(z)F_{\pm,\D}(z|\D_\phi)+\int_{\frac 12}^{1}dz\,  g_{\pm}(z)F_{\pm,\D}(z|\D_\phi)\,.
	\ee
	The functions $f$ and $g$ satisfy the conditions
	\be
	f_{\pm}(z)=\mp f_{\pm}(1-z)\,, \ \mathcal{R}_z f_{\pm}(z)=-g_{\pm}(z)\pm g_{\pm}(1-z)\,, \ g_{\pm}(z)=\e (1-z)^{2\D_\phi-2}f_{\pm}(\frac{z}{z-1})\,.
	\ee
	Here $\mathcal{R}_{z}$ denotes the real part, and $\e=+1(-1)$ for bosonic (fermionic) case. It is then possible to choose the kernel $f$ so that the functionals satisfy the orthogonality properties:
	\begin{align}
	\a_{\pm,n}^{B}(\D_m^{B})&=\d_{mn}\,,\hspace{1cm}\partial_\D\a_{\pm,n}^{B}(\D_m^{B})=-c^B_{\pm,n}(\D_\phi)\d_{m0}\,,\nonumber\\
	\b_{\pm,n}^{B}(\D_m^{B})&=0\,,\hspace{1.4cm}\partial_\D\b_{\pm,n}^{B}(\D_m^{B})=\d_{mn}-d_{\pm,n}^B(\D_\phi)\d_{m,0} \,.
	\end{align}
	Here we show the bosonic case with $\b^B_{\pm,0}=0$ and $c^B_{\pm,n},d^B_{\pm,n}$ are known. The fermionic functionals satisfy similar conditions replacing $B\to F$ and with $\b^F_{\pm,0}\ne 0$,  $c^F_{\pm,n}=d^F_{\pm,n}=0$\,. Explicit forms of all kernels can be found in Appendix A of \cite{Paulos:2019gtx}.
	
	A feature of the kernels that will be important in our discussion is the Regge limit $z\to i \infty$. These are as follows:
	\begin{align}\label{falloff}
	\omega_-:& \ \ f(z)\stackrel{z\to i \infty}{\sim} O(z^{-2})\nonumber\\
	\omega_+:& \ \ f(z)\stackrel{z\to i \infty}{\sim} O(z^{-3})\,.
	\end{align}
	This feature implies that $\omega_{-}$ ($\omega_+$) functionals can bootstrap (anti)crossing solutions with an $O(z^0)$ ($O(z)$) large $z$ behavior.
	
	It has been established that `$-$ type' analytic functionals are related to 1d Polyakov Bootstrap in an interesting way. The latter is the idea that a crossing symmetric correlator can be expanded in terms of manifestly crossing symmetric functions, called Polyakov blocks, in the following way
	\be\label{eq:Polboot}
	\mathcal{G}_-(z)=\sum_\D a_\D \mathcal{P}^{B,F}_{-,\D}(z)=\sum_\D a_\D G_{\D}(z)\,.
	\ee
	The equality on the right is the usual OPE. The Polyakov block $\mathcal{P}_{-,\D}(z)$ is given by a crossing symmetric sum of 4-point  Witten diagrams
	\be
	\mathcal{P}^{B,F}_{-,\D}(z)=W_{\D,0}^{(s),B,F}(z)+W_{\D,0}^{(t),B,F}(z)+W_{\D,0}^{(u),B,F}(z)+\k_-^{B,F}\, \mathcal{C}_{-}(z)\,.
	\ee
	Here $W_{\D,0}^{(i),B,F}(z)$ is a Witten exchange diagram (spin $0$ and dimension $\D$ exchange in $i=s,t,u$ channels) with bosonic/fermionic legs and drawn with a suitable choice of vertex. The $\mathcal{C}_{-}(z)$ is a crossing symmetric 4-point contact diagram. We require the Polyakov block to be Regge bounded i.e. $\mathcal{P}_{-,\D}(z)< \infty$ as $z\to i \infty$. This allows a single contact diagram with bosonic external legs with $\Phi^4$ vertex.  We define the Witten diagrams in Appendix \ref{app:Witten}.
	
	The Witten diagrams above can be decomposed into conformal blocks of dimensions $\D_n^{B,F}$. We can choose Polyakov blocks such that their block decomposition computes  functional actions (dropping $\D_\phi$ dependence for convenience) as follows:
	\be
	\mathcal{P}^{B,F}_{-,\D}(z)=G_{\D}(z)-\sum_n\Big[\a^{B,F}_{-,n}(\D)G_{\D^{B,F}_n}(z)+\b^{B,F}_{-,n}(\D)\partial_{\D}G_{\D_n^{B,F}}(z)\Big]\,.
	\ee
	So the statement of Polyakov Bootstrap \eqref{eq:Polboot} is identical to the functional bootstrap equations. While for the fermionic case this is automatic ($\k_-^F=0$), for the bosonic case one can choose $\k_-^B$ such that $\b^B_{-,0}=0$ and have the above correspondence. Notice that the basis \eqref{eq:basis} is nothing but the crossing symmetry equation for $\mathcal{P}_{-,\D}(z)$\,. For more details of Polyakov bootstrap in 1d see \cite{Mazac:2018ycv,Ghosh:2021ruh}.
	
	\subsection{The `+ type' Polyakov blocks}\label{Polyboot}
	
	An obvious question that one can now ask is if there exists an analogue of Polyakov blocks that computes `+ type' functional actions. Indeed there is such a function which we call a `$+$ type Polyakov block'. In this section we show how to build it from familiar Witten diagrams. 
	
	We introduce the following  notation for convenience:
	\begin{align}
	\D_{n,\ell}=2\D_\phi+2n+\ell\,.
	\end{align}
	This  is for consistency with the following sections. Note that we have $\D_{n,0}=\D^B_n$ and $\D_{n,1}=\D^F_{n}$\,.
	
	\subsubsection{Fermionic case}\label{sec:fermioniccase}
	Let us first discuss the case of fermionic functionals. Since $+$ type functionals act on  crossing antisymmetric vectors, the  $+$ type Polyakov blocks must be crossing antisymmetric. 
	Consider spin 1 exchange Witten diagrams in AdS$_2$ with \emph{bosonic} external legs. They have the conformal block decomposition (for this subsection we write $W_{\D,1}^{(i),B}\to W_{\D,1}^{(i)}$):
	\be\label{eq:decomp}
	W_{\D,1}^{(s)}=G_{\D}(z)+\sum_{n} \big( a_{n,1}^{(s)} G_{\D_{n,1}}(z)+ b_{n,1}^{(s)} \partial_{\D} G_{\D_{n,1}}(z)\big)\,.
	\ee
	In the crossed channels we have 
	\begin{align}\label{eq:decomp2}
	W_{\D,1}^{(-)}(z)&=\sum_{n} \big( a_{n,1}^{(t)} G_{\D_{n,0}}(z)+ b_{n,1}^{(t)} \partial_{\D} G_{\D_{n,0}}(z)\big)\,,\nonumber\\
	W_{\D,1}^{(+)}(z)&=\sum_{n} \big( \bar a_{n,1}^{(t)} G_{\D_{n,1}}(z)+ \bar b_{n,1}^{(t)} \partial_{\D} G_{\D_{n,1}}(z)\big)\,.
	\end{align}
	Here we defined $W_{\D,1}^{(\pm)}=\frac 12\big(W_{\D,1}^{(t)}\pm W_{\D,1}^{(u)}\big)$\,. The expressions for  $W^{(s)}_{\D,1}$  and other useful details are given Appendix \ref{app:Witten} (see \cite{Ghosh:2021ruh,Gopakumar:2018xqi,Zhou:2018sfz} for more on the conformal block decomposition).
	The crossed channel diagrams are related to the $s$-channel via:
	\be
	W_{\D,1}^{(t)}(z)=W_{\D,1}^{(s)}(1-z)\,, \hspace{1cm} W_{\D,1}^{(u)}(z)=(1-z)^{-2\D_\phi}\text{Re}W_{\D,1}^{(s)}\big(\frac{1}{1-z}\big)\,.
	\ee
	We also have
	\be
	W_{\D,1}^{(u)}(1-z)=-W_{\D,1}^{(u)}(z)\,.
	\ee

	A natural crossing antisymmetric Polyakov block is then 
	\begin{align}\label{Poly-fermion}
	\mathcal{P}_{+,\D}^F(z)&=W_{\D,1}^{(s)}(z)-W_{\D,1}^{(t)}(z)-W_{\D,1}^{(u)}(z)\,.\nonumber \\
	&=W_{\D,1}^{(s)}(z)-2W_{\D,1}^{(+)}(z)\,.
	\end{align}
	Notice that this object is totally crossing antisymmetric i.e. $\mathcal{P}_{+}(z)\to -\mathcal{P}_{+}(z)$ under $x_1\leftrightarrow x_3$ or $x_1\leftrightarrow x_4$. It is also antisymmetric in $x_1\leftrightarrow x_2$ which is not obvious from \eqref{Poly-fermion} although it is clear from their Mellin amplitudes (see Appendix \ref{app:Witten} and the next section).\footnote{All $x_i\leftrightarrow x_j$ are equivalent to simple transformations of Mellin variables, e.g.: $x_1\leftrightarrow x_3$ is equivalent to $s\leftrightarrow t+\D_\phi$.}
 	The anticrossing property implies the following:
	\begin{align}\label{eq:Polferm}
	&\mathcal{P}_{+\D}(z)=G_{\D}(z)- \sum_n\Big[\a^{F}_{+,n}(\D)G_{\D_{n,1}}(z)+\b^{F}_{+,n}(\D)\partial_{\D}G_{\D_{n,1}}(z)\Big]\nonumber\\
	\implies &F_{+,\D}(z)=\sum_{n}\Big[\a^F_{+,n}(\D|\D_\phi)F_{+,\D_{n,1}}(z)+\b^F_{+,n}(\D|\D_\phi)\partial_{\D}F_{+,\D_{n,1}}(z)\Big]\,,
	\end{align}
	where the coefficients are given by
	\begin{align}\label{funcaction}
	\a^F_{+,n}(\D|\D_\phi)&=-a_{n,1}^{(s)}+2\bar a_{n,1}^{(t)}\,,\nonumber\\
	\b^F_{+,n}(\D|\D_\phi)&=-b_{n,1}^{(s)}+2\bar b_{n,1}^{(t)}\,.
	\end{align}
	The coefficients $\a^F_{+,n}(\D|\D_\phi)$ and $\b^F_{+,n}(\D|\D_\phi)$ are named as such since they are equivalent to the + type fermionic functional actions. It can be verified by explicitly evaluating the actions  from \eqref{funcdef} that they match the expressions \eqref{funcaction}.
	
	We now point out an interesting feature. For the $-$ type bosonic functionals the missing $\b_{-,0}^B$ functional is attributed to the ambiguity of addition by a $\Phi^4$ contact Witten diagram to any crossing solution  that preserves Regge boundedness \cite{Mazac:2018ycv}. Similarly for the fermionic case there is no such ambiguity, and hence all $\omega_{-,n}^F$ are present. 
	Now recall from \eqref{falloff} that $\omega_+$ functionals can bootstrap anti-crossing symmetric solutions with a fall-off $O(z)$\,. It turns out that there is no anticrossing  `contact' diagram i.e. a crossing antisymmetric solution which decomposes into blocks and their derivatives of dimenisions $\D_{n,1}$ (only), and also has the Regge fall-off $O(z)$.\footnote{In section \ref{sec:contacts} we will see that the anticrossing contact diagram with the strictest Regge behavior is $\stackrel{s\to \infty}{\sim}O(s^3)$ in Mellin representation. Typically for an $s^{2j}$ behavior in Mellin space we have in position space 
	\be
\mathcal{G}(z)\stackrel{z\to i \infty}{\sim}z^{2j-1}\,.
\ee
This can be proved by assuming a vertex of the form $(\partial^j\Phi)^4$ in AdS$_2$ and computing a 4-point contact diagram \cite{Cornalba:2007fs,Costa:2012cb}.
}
	 This is consistent with the fact that fermionic $+$ type functionals $\omega_{+,n}^F$ exist for all $n\ge 0$. If there was such an anticrossing contact diagram, we would have had to subtract it from \eqref{Poly-fermion} to set the coefficient of one of the blocks to zero.

		\subsubsection{ Bosonic case}\label{sec:bosonic}

	Now let us consider the other case, bosonic functionals. One can guess that they require Witten diagrams with fermionic external legs.  In \cite{Faller:2017hyt}  the general $d$ fermionic Witten diagrams with a scalar exchange was computed (see also \cite{Binder:2020raz}). The $s$-channel scalar exchange Witten diagram in the AdS$_2$ case is given in terms of the bosonic Witten diagram as follows 
	\be
	W^{(s),F}_{\D,0}(z|\D_\phi)=z W^{(s),B}_{\D,0}\big(z|\D_\phi+\frac 12\big)\,.
	\ee
	The factor of $z$ is to account for the $z^{-2\D_\phi}$ included in our definitions. In a similar way let us define the following object:
	\be
	W^{(s),F}_{\D,1}(z|\D_\phi)=\frac{1}{z}W^{(s),B}_{\D,1}\big(z|\D_\phi-\frac 12\big)\,.
	\ee
	We will refer to this as the `fermionic Witten diagram' with an $s$-channel spin 1 exchange. Note that computing the Witten diagram with four  boundary-bulk fermionic propagators and bulk-bulk spin 1 propagator is  a non-trivial task which we do not perform. However it should be related to the above object by addition of a suitable contact diagram. The diagram we defined above already has everything we need for our discussion. 
	
	If the fermionic Witten $s$-channel diagram is defined as above, the corresponding crossed channel diagrams would be given by 
	\be
	W_{\D,1}^{(t),F}(z|\D_\phi)=W_{\D,1}^{(s),F}(1-z)\,, \hspace{1cm} W_{\D,1}^{(u),F}(z|\D_\phi)= (1-z)^{-2\D_\phi}\text{Re}W_{\D,1}^{(s),F}\big(\frac{1}{1-z}\big)\,.
	\ee
	Once again we have
	\be
	W_{\D,1}^{(u),F}(1-z|\D_\phi)=-W_{\D,1}^{(u),F}(z|\D_\phi)\,.
	\ee
	The block decomposition of the above objects are as follows:
	\begin{align}
	W_{\D,1}^{(s),F}(z)&=G_{\D}(z)+\sum_{n} \big( a_{n,1}^{(s),F} G_{\D_{n,0}}(z)+ b_{n,1}^{(s),F} \partial_{\D} G_{\D_{n,0}}(z)\big)\,.\nonumber\\
	W_{\D,1}^{(-),F}(z)&=\sum_{n} \big( a_{n,1}^{(t),F} G_{\D_{n,1}}(z)+ b_{n,1}^{(t),F} \partial_{\D} G_{\D_{n,1}}(z)\big)\,,\nonumber\\
	W_{\D,1}^{(+),F}(z)&=\sum_{n} \big( \bar a_{n,1}^{(t),F} G_{\D_{n,0}}(z)+ \bar b_{n,1}^{(t),F} \partial_{\D} G_{\D_{n,0}}(z)\big)\,,
	\end{align}
	where $W_{\D,1}^{(\pm),F}=\frac 12\big(W_{\D,1}^{(t),F}\pm W_{\D,1}^{(u),F}\big)$\,.
	
	With this we define an anticrossing object as follows:
	\begin{align}
	\mathcal{P}_{+}^{\text{exc}}(z)&=W_{\D,1}^{(s),F}(z)-W_{\D,1}^{(t),F}(z)-W_{\D,1}^{(u),F}(z)\,.\nonumber \\
	&=W_{\D,1}^{(s),F}(z)-2W_{\D,1}^{(+),F}(z)\,.
	\end{align}
	The `exc' stands for `exchange'. Under block decompositon $\mathcal{P}_{+}^{\text{exc}}$ gives a $G_{\D}(z)$ and $G_{\D_{n,0}}(z)$ $\forall n\in \mathbb{Z}_{+,0}$\,.
	
	However $\mathcal{P}_{+}^{\text{exc}}$  is not the Polyakov block yet. In fact there exists another  antisymmetric object which has a similar property. It is the analogue of a contact diagram, and is given by
	\be\label{eq:Pcon}
	\mathcal{P}_{+}^{\text{Con}}(z)=\mathcal{C}_+^{(s),F}(z)-\mathcal{C}_+^{(t),F}(z)-\mathcal{C}_+^{(u),F}(z)\,.
	\ee
	Here the first term on r.h.s. can be written in terms of a general dimension Mellin representation as follows:
	\be
	\mathcal{C}_+^{(s),F}(z)=z^{-2\D_\phi}\Bigg[\int_{-i\infty}^{i\infty} ds\,  dt\, z^{2s} \, (1-z)^{2t} \Gamma^2(\D_\phi+\frac 12-s)\Gamma^2(-t)\Gamma^2(s+t) \big(s+2t\big)\Bigg]\,.
	\ee
	The other two terms (crossed channels) are given by \footnote{Here $\mathcal{C}_+^{(t),F}$  and $\mathcal{C}_+^{(u),F}$ can be obtained by replacing  $\D_\phi+\frac 12-s\leftrightarrow -t$ and $\D_\phi+\frac 12-s\leftrightarrow s+t$ in the Mellin integral respectively. Rescaling the prefactor leads to the respective extra factors $\frac{1-z}{z}$ and $\frac{1}{z}$ in the Mellin amplitudes.}: 
	\be
	\mathcal{C}_+^{(t),F}(z)=\mathcal{C}_+^{(s),F}(1-z)\,, \hspace{1cm} \mathcal{C}_+^{(u),F}(z)=(1-z)^{-2\D_\phi}\text{Re}\, \mathcal{C}_+^{(s),F}\big(\frac{1}{1-z}\big)\,.
	\ee
	It is easy to verify that 
	\be
	\mathcal{C}_+^{(u),F}(1-z)=-\mathcal{C}_+^{(u),F}(z)\,.
	\ee
	Furthermore we have the block decomposition
	\begin{align}
	\mathcal{C}_+^{(s),F}(z)&=\sum_{n} \big( a_{n,C}^{(s),F} G_{\D_{n,0}}(z)+ b_{n,C}^{(s)} \partial_{\D} G_{\D_{n,0}}(z)\big)\,.\nonumber\\
	\mathcal{C}_+^{(-),F}(z)&=\sum_{n} \big( a_{n,C}^{(t),F} G_{\D_{n,1}}(z)+ b_{n,C}^{(t)} \partial_{\D} G_{\D_{n,1}}(z)\big)\,,\nonumber\\
	\mathcal{C}_+^{(+),F}(z)&=\sum_{n} \big( \bar a_{n,C}^{(t),F} G_{\D_{n,0}}(z)+ \bar b_{n,C}^{(t)} \partial_{\D} G_{\D_{n,0}}(z)\big)\,,
	\end{align}
	where $\mathcal{C}_+^{(\pm),F}=\frac 12\big(\mathcal{C}_+^{(t),F}\pm \mathcal{C}_+^{(u),F}\big)$\,.
	
	At large $z$ the anticrossing contact diagram $\mathcal{P}_+^{\text{Con}}(z)\sim O(z)$. So any solution obtained from analytic $+$ type bosonic functionals can be deformed by it. This is an ambiguity that should be taken into account in order to have the correct Polyakov block. Under block decomposition the Polyakov block must have one coefficient zero, so to reflect this ambiguity in the sum rules i.e. functional actions (see the last paragraph of previous subsection).
	
	Hence we define the \emph{bosonic} anti-crossing symmetric Polyakov block as follows 
	\begin{align}\label{Polybos}
	\mathcal{P}_{+}(z)&=\mathcal{P}_{+}^{\text{exc}}(z)+k\, \mathcal{P}_{+}^{\text{Con}}(z)\,\nonumber\\
	&=G_{\D}(z)- \sum_n\Big[\a^{B}_{+,n}(\D)G_{\D_{n,0}}(z)+\b^{B}_{+,n}(\D)\partial_{\D}G_{\D_{n,0}}(z)\Big]\,,
	\end{align}
	which implies the following relation 
	\be
	F_{+,\D}(z)=\sum_{n}\Big[\a_{+,n}^B(\D|\D_\phi)F_{+,\D_{n,0}}(z)+\b_{+,n}^B(\D|\D_\phi)\partial_{\D}F_{+,\D_{n,0}}(z)\Big]\,.
	\ee
	The constant $k$ in \eqref{Polybos} is chosen such that $\b_{+,0}^B(\D|\D_\phi)=0$. This ensures that all the coefficients $\a_{+,n}^B(\D|\D_\phi)$ and $\b_{+,n}^B(\D|\D_\phi)$ are actions of the bosonic + type functionals $\a_{+,n}^B$ and $\b_{+,n}^B$ respectively. This can be easily verified.

To summarize this section we have shown the existence of two  + type Polyakov blocks, `bosonic' and `fermionic'. They are built from  known Witten diagrams in AdS$_2$.  Through their conformal block decomposition shown in \eqref{eq:Polferm} and \eqref{Polybos} they compute the actions of + type analytic functionals.

	\section{General $d$  Polyakov blocks and dispersion relation}\label{sec:gendPolyblock}

	In this section we inititate a new formulation in general dimension to analyse crossing antisymmetric functions. This is analogous to the usual Polyakov Bootstrap for the crossing symmetric case. We propose a crossing antisymmetric basis built from  Witten diagrams. This is then explicitly demonstrated by setting up a dispersion relation for Mellin amplitudes of the functions. Most of the discussions in this section is formulated in Mellin space.
	
\begin{figure}[t]
		\begin{center}
			\includegraphics[width=0.9\textwidth]{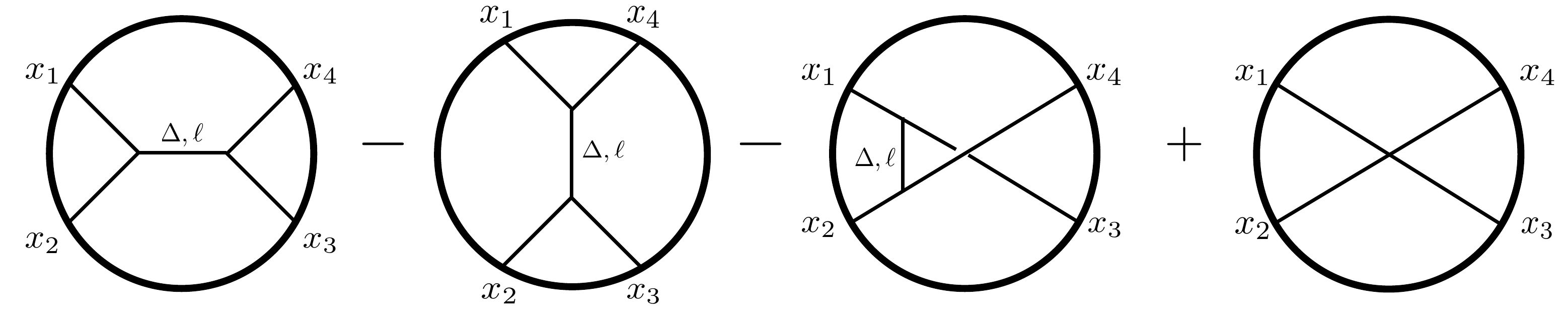}
		\end{center}
		\caption{
			The Polyakov block $\mathcal{P}_{+,\Delta,\ell}$ is a crossing antisymmetric combination of Witten exchange diagrams $W_{\D,\ell}(u,v)$ (with odd spin $\ell$) and antisymmetric `contact' diagrams' $\mathcal{C}_{+}(u,v)$. 
			\label{fig:Polyblock}}
	\end{figure}

	Let us define a new Polyakov block $\mathcal{P}_{+,\D,\ell}(u,v)$ in terms of which an anticrossing correlator can be expanded. We would like to have $\mathcal{P}_{+,\D,\ell}(u,v)$ to be a completely crossing antisymmetric object. By this we mean the following:
	\be\label{eq:anticrossing}
	\mathcal{P}_{+,\D,\ell}(u,v)=-\mathcal{P}_{+,\D,\ell}(v,u)=-u^{-\D_\phi}\mathcal{P}_{+,\D,\ell}(1/u,v/u)=-v^{-\D_\phi}\mathcal{P}_{+,\D,\ell}(u/v,1/v)\,.
	\ee
	It is possible to write such an object in terms of Witten exchange diagrams $W_{\D,\ell}(u,v)$ for odd $\ell$ in the following way (see Figure \ref{fig:Polyblock})
	\be\label{eq:Polydef}
	\mathcal{P}_{+,\D,\ell}(u,v)=W_{\D,\ell}^{(s)}(u,v)-W_{\D,\ell}^{(t)}(u,v)-W_{\D,\ell}^{(u)}(u,v)+\text{contacts}\,.
	\ee
	These diagrams are all defined in AdS$_{d+1}$ with identical external scalars of dimension $\D_\phi$\,. 
	It is convenient to write them in the Mellin representation as follows:
	\be\label{Mellinrep}
	W_{\D,\ell}(u,v) =\int ds_1 ds_2 \,  u^{s_1+\frac{2\D_\phi}{3}}v^{s_2-\frac{\D_\phi}{3}}\Bigg[\prod_{i=1}^{3}\Gamma^2\big(\frac{\D_\phi}{3}-s_i\big)\Bigg] M_{\D,\ell}(s_1,s_2)\,,
	\ee
	where $s_3=-s_1-s_2$. The $M_{\D,\ell}(s_1,s_2)$ is a Mellin amplitude defined in Appendix \ref{app:Witten}. The `contacts' denote crossing antisymmetric functions $\mathcal{C}_+(u,v)$ with polynomial Mellin amplitudes which we discuss in a moment.  We follow the conventions of \cite{Gopakumar:2021dvg,Sinha:2020win}. The $s_i$ are related to the $s,t,u$ variables of previous section by the simple shifts $s_1=s-\frac{2 \Delta _{\phi }}{3}$ and $s_2=\frac{\Delta _{\phi }}{3}+t$.

	The totally crossing antisymmetric property \eqref{eq:anticrossing} directly follows from the odd spin Mack polynomials in the Mellin amplitudes, and is easily seen from the meromorphic form \eqref{eq:mero}.
	The exchange Witten diagrams  have the following conformal block decomposition: 
	\begin{align}
	W_{\D,\ell}^{(s)}(u,v)&=G_{\D,\ell}(z)+\sum_{n} \big[ a_{n,\ell}^{(s)}(\D|\D_\phi) G_{\D_{n,\ell},\ell}(u,v)+ b_{n,\ell}^{(s)}(\D|\D_\phi) \partial_{\D} G_{\D_{n,\ell},\ell}(u,v)\big]\,,\nonumber\\
	W_{\D,\ell}^{(-)}(u,v)&=\sum_{n} \sum_{\ell' \text{even}} \big[ a_{n,\ell'|\ell}^{(t)}(\D|\D_\phi) G_{\D_{n,\ell'},\ell}(u,v)+ b_{n,\ell'|\ell}^{(t)}(\D|\D_\phi) \partial_{\D} G_{\D_{n,\ell'},\ell'}(u,v)\big]\,,\nonumber\\
	W_{\D,\ell}^{(+)}(u,v)&=\sum_{n} \sum_{\ell' \text{odd}} \big[ \bar a_{n,\ell'|\ell}^{(t)}(\D|\D_\phi) G_{\D_{n,\ell'},\ell'}(u,v)+ \bar b_{n,\ell'|\ell}^{(t)}(\D|\D_\phi) \partial_{\D} G_{\D_{n,\ell'},\ell'}(u,v)\big]\,.
	\end{align}
	Here $W_{\D,\ell}^{(\pm)}=\frac 12\big(W_{\D,\ell}^{(t)}\pm W_{\D,\ell}^{(u)}\big)$ and $\D_{n,\ell}=2\D_\phi+2n+\ell$\,. 
	
	Then antisymmetry of \eqref{eq:Polydef} implies the following  equation:
	\be
	F_{+,\D,\ell}(u,v)=\sum_{n}\sum_{\ell' \text{odd}}\Big[\a_{n,\ell|\ell'}F_{+,\D_{n,\ell'},\ell}(u,v)+\b_{n,\ell|\ell'}\partial_{\D}F_{+,\D_{n,\ell'},\ell}(u,v)\Big]\,.
	\ee
	The $\a$ and $\b$ are given by
	\begin{align}
	\a_{n,\ell'|\ell}(\D|\D_\phi)&=-a_{n,\ell}^{(s)}(\D|\D_\phi)\d_{\ell\ell'}+2\bar a_{n,\ell'|\ell}^{(t)}(\D|\D_\phi)+\cdots\,,\label{eq:funcactiondef1}\\
	\b_{n,\ell'|\ell}(\D|\D_\phi)&=-b_{n,\ell}^{(s)}(\D|\D_\phi)\d_{\ell\ell'}+2\bar b_{n,\ell'|\ell}^{(t)}(\D|\D_\phi)+\cdots\label{eq:funcactiondef2}\,.
	\end{align}
	Here $\cdots$ are contact diagram contributions discussed below.
	
	Similar to the usual crossing symmetric case this implies a basis of expansion for $F_{+,\D,\ell}(u,v)$. If the anticrossing correlator has a conformal block decomposition in  $G_{\D,\ell}(u,v)$ like \eqref{confblockdeco} we should be able to write the equations:
	\be\label{eq:Polyeqn}
	\sum_{\D,\ell}a_{\D,\ell}\a_{n,\ell|\ell'}(\D|\D_\phi)=\sum_{\D,\ell}a_{\D,\ell}\b_{n,\ell|\ell'}(\D|\D_\phi)=0\,.
	\ee
	These are analogous to the `Polyakov Bootstrap equations' or `Polyakov conditions' in the usual crossing symmetric case. They are equivalent to a set of functionals for crossing antisymmetry equations just like its 1d counterpart, the $+$ type (fermionic) analytic functional of section \ref{sec:fermioniccase}. We return to these conditions in a moment.
	
	\subsection{Contact diagrams}\label{sec:contacts}
	We now address the missing piece of this formulation: the contact diagrams. 
	For a contact diagram $\mathcal{C}(u,v)$ the corresponding Mellin amplitude $M_{\mathcal{C}}(s_1,s_2)$ is a polynomial. A completely antisymmetric polynomial Mellin amplitude may be written in terms of a crosing symetric one in the following way:
	\be\label{eq:Mellincont}
	M_{\mathcal{C},+}(s_1,s_2)=(s_1-s_2)(s_2-s_3)(s_3-s_1)M_{\mathcal{C},-}(s_1,s_2)\,.
	\ee
	The superscripts `$+$' and `$-$' respectively denote crossing symmetric and antisymmetric as usual.  It is known that $M_{\mathcal{C},-}$ are polynomials of the invariants $x=-(s_1 s_2+s_2 s_3+s_3 s_1)$ and $y=-s_1 s_2 s_3$\,. For convenience let us also denote $w=(s_1-s_2)(s_2-s_3)(s_3-s_1)$. So we get  
	\be\label{eq:contact}
	M_{\mathcal{C},+}(s_1,s_2)=w\sum_{p,q} c_{p,q}x^p y^q\,.
	\ee
	Note that $M_{\mathcal{C},+}(s_1,s_2)$ can have the lowest degree 3 in the variables. It is followed by $5,6,$ etc\,.
	
	We may now define the Polyakov block as follows:
	\be\label{eq:Polydef2}
	\mathcal{P}_{+,\D,\ell}(u,v)=W_{\D,\ell}^{(s)}(u,v)-W_{\D,\ell}^{(t)}(u,v)-W_{\D,\ell}^{(u)}(u,v)+\mathcal{C}_{+}(u,v)(u,v)\,.
	\ee
	The contact diagram $\mathcal{C}_+(u,v)$ has the conformal block decomposition:
	\be
	\mathcal{C}_+(u,v)=\sum_{n}\sum_{\ell' \text{odd}} \big[ a_{n,\ell'|\mathcal{C}}(\D_\phi) G_{\D_{n,\ell'}}(u,v)+ b_{n,\ell'|\mathcal{C}}(\D_\phi) \partial_{\D} G_{\D_{n,\ell'}}(u,v)\big]\,.
	\ee
	It is straightforward to obtain the coefficients (see Appendix \ref{app:Witten})\,. The expressions \eqref{eq:funcactiondef1} and \eqref{eq:funcactiondef2} are now modified with the additional terms $a_{n,\ell'|\mathcal{C}}$ and $ b_{n,\ell'|\mathcal{C}}$ respectively.

	\subsection{Dispersion relation}
	
	We would like to show that an expansion of an anticrossing correlator in $\mathcal{P}_{+,\D,\ell}$ is indeed valid, thereby 
	fixing the coefficients $c_{p,q}$\,. For this we follow the framework set up for the crossing symmetric case in \cite{Gopakumar:2021dvg,Sinha:2020win}. The idea in those papers was to write down a dispersion relation for the correlator, which is manifestly crossing symmetric. This leads to introduction of new spurious poles in the Mellin amplitude, that are called `nonlocal' terms. Subtracting out these terms one is left with crossing symmetric sums of exchange Witten diagrams and polynomial terms. The latter fixes the required contact term ambiguities. We review the details briefly in Appendix  \ref{App:dispersion}.
	
	Consider the Mellin amplitude $\mathcal{M}_+(s_1,s_2)$ of a crossing antisymmetric function. We will drop the `$+$' subscript below since we consider only the antisymmetric case in this subsection. We will assume the usual Regge bounded condition \footnote{Interestingly the final dispersion formula \eqref{eq:antidisp} allows a more relaxed Regge behavior of $O(s_1^{3-\e})$. Also we will see that the associated sum rules will reduce in 1d to the ones of analytic functionals discussed in section \ref{sec:1dfunc}. Indeed  the allowed Regge behavior is, quite nicely, the same with both methods.}
	\be
	\mathcal{M}(s_1,s_2)\stackrel{s_1\to \infty}{\sim} O(s_1^{2-\e})\, \ \text{with} \ \e> 0\,, \text{ $s_2$ fixed}\,.
	\ee
	We will write down a twice subtracted dispersion relation for this in a manifestly crossing antisymmetric way, following a strategy  similar  to \cite{Gopakumar:2021dvg,Sinha:2020win}. To write such a relation we first use a change of variables:
	\be
	s_i=a-\frac{a(z-z_i)^3}{z^3-1}\,,
	\ee
	where $z_i$ denotes cube roots of 1, and 
	\be
	a=\frac{s_1 s_2 s_3}{s_1 s_2 +s_2 s_3 +s_3 s_1}=\frac{y}{x}\,.
	\ee
	The variable $a$ is crossing symmetric. \footnote{The reader should not confuse this $z$ with the cross ratios. In this section we use only $u,v$ in position space.} The `physical cuts'  (analogous to QFT  scattering amplitudes) in each $s_i$ plane can be mapped to the $z$-plane. 
	Instead of the usual technique of integrating along $s_1$ keeping $s_2$ fixed, our dispersion relation is in the $z$ variable keeping  $a$ fixed. 
	We impose the condition of  antisymmetry on the discontinuity of the Mellin amplitude accross the cut, and rewrite the relation in terms of the original variables $s_i$. We show the steps in detail in Appendix \ref{App:dispersion}. The result is the crossing antisymmetric dispersion relation:
	\be\label{eq:antidisp}
	\mathcal{M}(s_1,s_2)=\frac{1}{\pi}\int_{\tau^{(0)}}^{\infty}\frac{ds_1'}{s_1'}\mathcal{A}(s_1';s_2^+(s_1',a))H(s_1';s_1,s_2,s_3)\,.
	\ee
	In the above $\mathcal{A}(s_1,s_2)$ is the $s$-channel discontinuity. The integration is over the cut $s_1\ge \tau^{(0)}$ which for a CFT correlator is basically a series of poles starting at $\tau^{(0)}$. 
	Here the crossing antisymmetric kernel $H$ given by
	\be
	H(s;s_1,s_2,s_3)=\Big({\frac{s-a}{s+3a}}\Big)^{\frac 12}\Bigg[\frac{s_2-s_3}{s-s_1}+\frac{s_3-s_1}{s-s_2}+\frac{s_1-s_2}{s-s_3}\Bigg]\,.
	\ee
	Finally in the absorptive part we also have:
	\begin{align}
	s_2^+(s,a)&=-\frac{s}{2}\Big[{1-\sqrt{\frac{s+3a}{s-a}}}\Big]\,.
	\end{align}
	
	We now demand that the Mellin amplitude has an expansion in the crossing symmetric invariants $a$ and $x$ as follows
	\be
	\mathcal{M}(s_1,s_2)=w\sum_{p,q=0}^{\infty}\mathcal{M}_{p,q}x^{p+q}a^q\,,
	\ee
	Replacing $a=y/x$ we get an expansion of the form $x^py^q$.
For an arbitrary amplitude from \eqref{eq:antidisp} one can  get negative powers of $x$ i.e. $p<0$. These are unphysical (nonlocal) powers which we set to 0 by imposing the `locality constraints' :
	\be\label{eq:localcondn}
	\mathcal{M}_{p<0,q}=0\,.
	\ee
	The counterparts of these conditions in the crossing symmetric case were shown in \cite{Sinha:2020win} to be equivalent to the `null constraints' \cite{Caron-Huot:2020cmc,Tolley:2020gtv} that lead to two sided bounds on Wilson coefficients in effective field theories. The new conditions \eqref{eq:localcondn} are hence a set of independent constraints when the EFT scattering amplitude has an antisymmetric sector. However in this paper we only consider the CFT case.

	Now let us see how the dispersion relation reproduces the Polyakov block $\mathcal{P}_{\D,\ell}$. If the antisymmetric correlator decomposes in the $s$-channel into operators of dimension $\D$ and spin $\ell$  then the discontinuity $\mathcal{A}$ can be written as a sum over the `partial waves':
	\be\label{eq:branch}
	\mathcal{A}(s_1,s_2)=\pi\sum_{\D,\ell,k}c_{\D,\ell}^{(k)}P_{\D,\ell}(\tau_k,s_2)\d(\tau_k-s_1)\,.
	\ee
	Here we have $\tau_k=\frac{\D-\ell}{2}+k-\frac{2\D_\phi}{3}$ and $k\in \mathbb{Z}_{\ge 0}$\,. Also $P_{\D,\ell}(s_1,s_2)$ is a shifted  Mack polynomial, and $c^{(k)}_{\D,\ell}$ is the squared OPE coefficient $a_{\Delta,\ell}$ times a  normalization. We define them in  Appendix \ref{app:Witten}.
	
	Then we get from \eqref{eq:antidisp}
	\be\label{eq:Mellin}
	\mathcal{M}(s_1,s_2)=\sum_{\D,\ell,k}c_{\D,\ell}^{(k)}\mathcal{M}_{\D,\ell,k}(s_1,s_2)\,,
	\ee
	where 
	\be\label{eq:partialwv}
	\mathcal{M}_{\D,\ell,k}(s_1,s_2)=\frac{1}{\tau_k}Q_{\ell,k}^{(\D)}(a)H(\tau_k,s_1,s_2,s_3)\,,
	\ee
	and $Q_{\ell,k}^{(\D)}(a)=P_{\D,\ell}(\tau_k,s_2'(\tau_k,a))$\,.

	We would like to show that this reduces to a crossing antisymmetric combination of Witten diagrams when $\ell$ is odd. For this let us define the quantity
	\be\label{eq:Witten}
    \mathcal{M}_{\D,\ell,k}^{(0)}(s_1,s_2)=\frac{P_{\D,\ell}(s_1,s_2)}{s_1-\tau_k}-\frac{P_{\D,\ell}(s_2,s_1)}{s_2-\tau_k}-\frac{P_{\D,\ell}(s_3,s_2)}{s_3-\tau_k}\,,
    \ee	
	with $s_3=-s_1-s_2$\,. It can be checked for every odd $\ell$ that this is antisymmetric under $s_i\leftrightarrow s_j$ $(i\ne j)$. Each term in $\mathcal{M}_{\D,\ell,k}^{(0)}(s_1,s_2)$ is a pole of the Mellin amplitude of exchange Witten diagram $W_{\D,\ell}^{(i)}$ when we write it in its meromorphic form (see \eqref{eq:mero} in Appendix \ref{app:Witten}). 
	
	We now expand  $\mathcal{M}_{\D,\ell,k}$ and $\mathcal{M}^{(0)}_{\D,\ell,k}$ in small $x$ and $a$, and look at their difference.

	\begin{align}\label{eq:dispersionmatch}
	\mathcal{M}_{\D,\ell,k}(s_1,s_2)=&\mathcal{M}_{\D,\ell,k}^{(0)}(s_1,s_2)+w\sum_{p,q=0}^{2p+3q\le L}C_{p,q}^{(\ell)}x^{p+q}a^{q}\nonumber\\
	&+w\sum_{p<0}\widetilde{C}^{(\ell)}_{p,q}x^{p+q}a^q\,.
	\end{align}
The second term is precisely the polynomial pieces that define the contact terms we wanted to fix in \eqref{eq:Polydef2}. For each $\ell$ there is a finite number of such polynomials up to a maximum degree $L$ (note that $x^py^q$ is a $2p+3q$ degree polynomial).
	Finally the last term i.e. second line of r.h.s. vanishes when we collectively impose the locality constraints \eqref{eq:localcondn} on the full amplitude\,. 
	
	For $\ell=1$ we get $C_{p,q}^{(\ell)}=\widetilde{C}^{(\ell)}_{p,q}=0$\,. For $\ell=3$ we get $L=0$ and $C_{0,0}^{(\ell)}=\frac{1}{16\tau_k}$\,. For $\ell=5$ we have $L=4$ which we work out in Appendix \ref{app:ell5}. 
	
	To conclude, we have shown that if an anticrossing correlator decomposes into odd spins it can be expanded in $+$ type Polyakov blocks. For a more general anticrossing function  an expansion like \eqref{eq:Mellin} remains valid.

	\subsection{Polyakov Conditions}\label{sec:polycond}
	Let us discuss how to bootstrap the Mellin amplitude to obtain the OPE data. For this let us reinstate the subscripts ($+$)$-$ for crossing (anti)symmetric correlators. 
    In terms of $\mathcal{M}_{\pm}$  they read
	\be\label{Mellinrep1}
	\mathcal{G}_{\pm}(u,v) =\int ds_1 ds_2 \,  u^{s_1+\frac{2\D_\phi}{3}}v^{s_2-\frac{\D_\phi}{3}}\Bigg[\prod_{i=1}^{3}\Gamma^2\big(\frac{\D_\phi}{3}-s_i\big)\Bigg] \mathcal{M}_{\pm}(s_1,s_2)\,,
	\ee 
	Since the only poles in $s_i$ that contribute to OPE come from \eqref{eq:branch} the poles from the Gamma function measure are spurious.\footnote{We assume no operator at the exact locations $\D=2\D_\phi+2p+\ell$.} So at these locations the Mellin amplitude must have double zeroes. These are the
	 `Polyakov conditions'. We write them as follows:
	\begin{align}\label{eq:Polycond}
	\mathfrak{F}_{\pm,p}(s_2)&\equiv \mathcal{M}_\pm(s_1=\frac{\D_\phi}{3}+p,s_2)=0\,,\nonumber\\
	\widetilde{\mathfrak{F}}_{\pm,p}(s_2)&\equiv \partial_{s_1}\mathcal{M}_\pm(s_1=\frac{\D_\phi}{3}+p,s_2)=0\,,
	\end{align}
	for  all $p\in \mathbb{Z}_{\ge 0}$. Let us focus on the first set of equations i.e. $\mathfrak{F}_{\pm,p}(s_2)=0$. Using \eqref{eq:Mellin} and summing over $k$ we can write $\mathfrak{F}^{(r)}_{\pm,p}(s_2)=\sum_{\Delta,\ell}a_{\D,\ell} \, \mathfrak{F}^{(r)}_{\pm,p,\D,\ell}(s_2)$. 
	We may Taylor expand this around $s_2=0$ to get a set of infinite sum rules as follows:  \footnote{\label{footnote:Ploy}
		For the $-$ case it was shown in \cite{Gopakumar:2021dvg} that with a special combination 
		\begin{align}
		\Omega_{\pm,p_1,p_2,p_3}(s_2)=&-\Big[\frac{\mathfrak{F}_{\pm,p_1}}{(p_1-p_2)(p_1+p_3+s_2+\frac{\D_\phi}{3})}+(p_1\leftrightarrow p_2)\Big]
		-\frac{\mathfrak{F}_{\pm,p_3}}{(p_1+p_3+s_2+\frac{\D_\phi}{3})(p_2+p_3+s_2+\frac{\D_\phi}{3})}\,
		\end{align}
		the sum rules \eqref{eq:Polycond} are equivalent to those obtained from nonperturbative Mellin amplitudes \cite{Carmi:2020ekr}. 
	}
	\be\label{eq:Polycond2}
	\mathfrak{F}_{\pm,p,\D,\ell}(s_2)=\sum_{r=0}^{\infty}(s_2)^r \mathfrak{F}^{(r)}_{\pm,p,\D,\ell}\, \  \implies \ \sum_{\D,\ell}a_{\D,\ell}\mathfrak{F}^{(r)}_{\pm,p,\D,\ell}=0\,.
	\ee
	A similar set of sum rules are obtained also from $\widetilde{\mathfrak{F}}_{\pm,p}(s_2)$. Note that these sum rules are equivalent to the conditions \eqref{eq:Polyeqn} since they are related by a Mellin integral.

	We point out that the description of $\mathcal{M}_{-}$ ($\mathcal{M}_+$) in terms of Witten diagrams is not possible when $\ell$ is odd (even) as the crossing (anti)symmetric properties are not well-defined. However the dispersion relation representation  holds for any spin. In particular for the $-$ case it can be checked that they lead to identical sum rules as \cite{Carmi:2020ekr} for any spin (see footnote \ref{footnote:Ploy}).  This is useful e.g. in global symmetry problems where the (anti)crossing equation may involve all spins (see next section).
	
	\paragraph{Numerical checks:} We have numerically tested our proposals in two different examples. First with a fictitious Mellin amplitude, we have checked that an expansion in $x$ and $a$ indeed works and locality conditions \eqref{eq:localcondn} are satisfied. In the second example we have worked with an antisymmetric sector of a 2d Wess-Zumino-Witten (WZW) model. For this we have shown the working of Polyakov conditions \eqref{eq:Polycond}. Both analyses are discussed in Appendix \ref{app:numchecks}.

	\section{Applications to functionals}\label{sec:applications}
	
	The $+$ type Polyakov blocks can be useful in a number of Bootstrap applications. The most immediate examples are computing the `product functionals' in higher $d$, and bootstrapping CFTs with global symmetries. In this section we discuss them in turn. 
	
	\subsection{Product Functionals in even $d$}\label{sec:prodfunc}
	
We begin by reviewing the  product functionals that were introduced in \cite{Paulos:2019gtx} as a set of efficient functionals in even $d$. In 2d it was shown that they have nice positivity properties making them ideal for numerical applications. 
The main advantage of these functionals is that they are built of 1d analytic functionals $\omega_\pm$. Through the Polyakov blocks we now understand how to compute their action in a simple way. 

In this subsection we only focus on $d=2$. We review the cases of $d=4,6$ in Appendix \ref{app:productfunc}.  The 2d conformal blocks are given by:
		\be
	G^{d=2}_{\D,\ell}(z,\bar z)=\frac 12\Big[G_{\frac{\tau}{2}}(z|\text{\tiny $\frac{\D_\phi}{2}$})G_{\frac{\rho}{2}}(\zb|\text{\tiny $\frac{\D_\phi}{2}$})+(z\leftrightarrow \bar{z})\Big]\,, \ \ \tau=\D-\ell\,, \ \rho=\D+\ell\,.
	\ee
	The crossing vector is given by:
		\be
	F^{d=2}_{-,\D,\ell}(z,\bar z|\D_\phi)=\frac 14 \big[ F_{-,\frac{\tau}{2}}(z|\text{\tiny$\frac{\D_\phi}{2}$})F_{+,\frac{\rho}{2}}(\zb|\text{\tiny$\frac{\D_\phi}{2}$})+ F_{-,\frac{\rho}{2}}(z|\text{\tiny$\frac{\D_\phi}{2}$})F_{+,\frac{\tau}{2}}(\zb|\text{\tiny$\frac{\D_\phi}{2}$}) +(z\leftrightarrow \bar z)\big]\,.
	\ee
	Notice that the $z$ and $\zb$ dependencies have separated out. If we ensure that the $z$ dependence is only through $F_-$, and $\zb$ through $F_+$ then we can write a functional for $F^{d=2}_{-,\D,\ell}$ as products of $\omega_-$ and $\omega_+$. Therefore we choose the 2d functionals as follows:
	\be
	\omega^{(1)}_-\otimes\omega^{(2)}_+(\D,\ell)=2\int_{1}^{\infty}\frac{dz d{\bar z}}{\pi^2}h^{(1)}_-(z)h^{(2)}_+(\bar z)\big[\mathcal{I}_z\mathcal{I}_{\bar z}F_{-,\D,\ell}(z,\bar z)+\mathcal{I}_z\mathcal{I}_{\bar z}F_{-,\D,\ell}(z,1-\bar z)\big]\,.
	\ee
	The structure is symmetrized in $\bar z\to 1-\bar z$ to have the above mentioned feature. All kernels $h^{(1)}_-,h^{(2)}_+$ are chosen from 1d functional kernels such that we have the following four types of functionals (suppressing $B,F$ superscripts as any bosonic/fermionic functional should work):
	\begin{align}
	\omega^{(1)}_-\omega^{(2)}_+\in \Big\{\a_{-,n}\a_{+,m}\,, \ \a_{-,n}\b_{+,m}\,, \ \b_{-,n}\a_{+,m}\,, \  \b_{-,n}\b_{+,m}\Big\}\,.
	\end{align}
	We have simplified the notation $	\omega^{(1)}_-\omega^{(2)}_+ \equiv 	\omega^{(1)}_-\otimes\omega^{(2)}_+$\,.
	Each functional action is given by
	\be\label{eq:2daction}
	\omega^{(1)}_-\omega^{(2)}_+(\D,\ell)=\frac 12\big[\omega^{(1)}_-(\tau|\D_\phi|2)\omega^{(2)}_+(\rho|\D_\phi|2)+\omega^{(1)}_-(\rho|\D_\phi|2)\omega^{(2)}_+(\tau|\D_\phi|2)\big]\,.
	\ee
	Here we defined for convenience
	$\omega_{\pm}(\D|\D_\phi|2)=\omega_{\pm}(\frac{\D}{2}|\frac{\D_\phi}{2})$. Using the results from section \ref{Polyboot} these are entirely determined from conformal block decomposition of Witten diagrams.
	The $+$ sign in the combination \eqref{eq:2daction} ensures the positive behavior of product functional actions. This was used in \cite{Paulos:2019gtx} with $\beta_{-,0}^{F}\a_{+,0}^F$, for instance, to obtain an upper bound on the lowest non-identity operator. 
	
	Let us also explore the case of crossing antisymmetric vector $F^{d=2}_{+,\D,\ell}(z,\zb)$ which can be written as
	\be
	F^{d=2}_{+,\D,\ell}(z,\bar z|\D_\phi)=\frac 14 \big[ F_{-,\frac{\tau}{2}}(z|\text{\tiny$\frac{\D_\phi}{2}$})F_{-,\frac{\rho}{2}}(\zb|\text{\tiny$\frac{\D_\phi}{2}$})+ F_{+,\frac{\rho}{2}}(z|\text{\tiny$\frac{\D_\phi}{2}$})F_{+,\frac{\tau}{2}}(\zb|\text{\tiny$\frac{\D_\phi}{2}$}) +(z\leftrightarrow \bar z)\big]\,.\,.
	\ee
	As before we have to symmetrize/antisymmetrize in $z\to 1-z$ to have the necessary variable separation. So we choose the functionals as follows:
	\be
	\omega^{(1)}_{\pm}\omega^{(2)}_{\pm}(\D,\ell)=2\int_{1}^{\infty}\frac{dz d{\bar z}}{\pi^2}h^{(1)}_\pm(z)h^{(2)}_\pm(\bar z)\big[\mathcal{I}_z\mathcal{I}_{\bar z}F_{+,\D,\ell}(z,\bar z)\pm\mathcal{I}_z\mathcal{I}_{\bar z}F_{+,\D,\ell}(z,1-\bar z)\big]\,.
	\ee
	The crossing antisymmetric product functionals and their respective functional actions are as follows
	\begin{align}
	&\omega^{(1)}_{\pm}\omega^{(2)}_{\pm}  \in\Big\{\a_{\pm,n}\a_{\pm,m}\,, \ \a_{\pm,n}\b_{\pm,m}\,, \ \b_{\pm,n}\b_{\pm,m}\Big\}\,.\\ 
	&\omega^{(1)}_{\pm}\omega^{(2)}_{\pm}(\D,\ell)  =\omega^{(1)}_{\pm}(\tau|\D_\phi|2)\omega^{(2)}_{\pm}(\rho|\D_\phi|2)+\omega^{(2)}_{\pm}(\tau|\D_\phi|2)\omega^{(1)}_{\pm}(\rho|\D_\phi|2)\,.
	\end{align}

	\subsection{Global symmetry - Simple functionals}\label{sec:simple}
	\subsubsection{$d=1$}
	An important application for crossing antisymmetric functionals is in the problem of global symmetries.  
	Let us recall from section \ref{sec:intro} that if we write the correlator of charged scalar fields $\phi_i$ in terms of the components $\mathcal{G}^{\mf a}$ for each irrep $\mf a$ appearing in the OPE, then under crossing we get the relation:
	\be\label{crossing1}
	\mathcal{G}^{\mf a}(z)=C^{\mf a \mf b} \mathcal{G}^{\mf b}(1-z)\,,
	\ee
	where $C^{\mf b\mf a}$ is the crossing matrix. We can define the projector matrices $P^{\mf b\mf a}_{\pm}=\frac 12(\d^{\mf b\mf a}+C^{\mf b\mf a})$ in terms of which one may write the crossing vectors
	\be\label{globalvec}
	\mathcal{F}^{\mf b|\mf a}(z)=P^{\mf b \mf a}_{+}F_{-,\D}(z)+P^{\mf b \mf a}_{-}F_{+,\D}(z)\,.
	\ee
	With the above crossing vectors one can write the crossing equation \eqref{crossing1} as follows:
	\be\label{crossing2}
	\sum_{\mf b}\sum_{\D}a_{\D}^{\mf b}\mathcal{F}^{\mf b|\mf a}(z)=0\,.
	\ee
	In \cite{Ghosh:2021ruh} a set of global symmetry functionals, called `simple functionals' were introduced for the above problem. They have the following structure:
	\be\label{simpfunc}
	\omega^{\mathfrak{b}|\mathfrak{a}}=P^{\mf b \mf a}_{+}\omega_{-}+P^{\mf b \mf a}_{-}\omega_{+}\,.
	\ee
	Here $\omega_{\pm}\in \{\a_{\pm,n}^{B,F},\b_{\pm,n}^{B,F}\}$ i.e. the 1d functionals of uncharged case. In the above notation for $\omega^{\mathfrak{b}|\mathfrak{a}}$ the superscript  $\mf b$ denotes a label and $|\mf a$ a component.
	The action of these functionals on the crossing vectors  $\mathcal{F}^{\mf c|\mf a}$ \eqref{globalvec} is given by
	\be
	\omega^{\mf b}(\mf c,\D)=P^{\mf b\mf c}_{+}\omega_{-}(\D)+P^{\mf b\mf c}_{-}\omega_{+}(\D)\,.
	\ee
	The action of these functionals on crossing symmetry \eqref{crossing2} are entirely determined by our knowledge of the $\omega_{\pm}(\D)$, which are given by Witten diagram decomposition coefficients (section \ref{Polyboot}). For numerical applications simple functionals are useful as they retain the positivity properties of the uncharged functionals. For instance any bound obtained using 1d functionals (say $\omega_{-}$) can be easily generalized to global symmetries if one can write a linear combination of $\omega^{\mf b}$ whose action is proportional to only $\omega_{-}$ with positive coefficients.
	
	\subsubsection{$d=2$ with product functionals}\label{sec:productfunc2d}
	
	The idea of simple functionals is not restricted to 1d. Indeed for any $d$ one can consider \eqref{simpfunc} choosing $\omega_{\pm}$ as appropriate in that dimension. In 2d we may choose them to be the product functionals 
	\be
	\omega_{-}\to \omega_{-}^{(1)}\omega_{+}^{(2)}\,, \ \ \omega_{+}\to \omega_{\pm}^{(3)}\omega_{\pm}^{(4)}\,.
	\ee
	The action of 2d simple product functionals is given by
	\be
	\omega^{\mf b}(\mf c,\D,\ell)=P_{+}^{\mf b\mf c}\big[\omega_{-}^{(1)}\omega_{+}^{(2)}(\D,\ell)\big]+P_{-}^{\mf b\mf c}\big[\omega_{\pm}^{(3)}\omega_{\pm}^{(4)}(\D,\ell)\big]\,.
	\ee
	This allows us to translate any  bound found with product functionals in CFTs without global symmetry to bounds for global symmetry problems. A simple demonstration is done in Appendix \ref{sec:bounds} where we extend an analytic bound for OPE density for 2d uncharged CFTs  to the charged case. As mentioned in section \ref{sec:prodfunc} the positivity properties of product functionals can be useful to obtain sharp numerical bounds for global symmetry problems.  
	
	Of course the $d=2$ simple product functionals case can easily be generalized to $d=4$ and $6$ (see Appendix \ref{app:productfunc}).

	\subsubsection{General $d$}
	To obtain simple functionals for general dimensions we may use the Polyakov condition sum rules \eqref{sec:polycond}. In this case we may use $\omega_\pm \to \mathfrak{F}_{\pm,p}, \widetilde{\mathfrak{F}}_{\pm,p}$ so that simple functional actions are given by
	\be
	\omega^{\mf b}({\mf c,\Delta,\ell})\in \big\{P_{+}^{\mf b\mf c}\mathfrak{F}^{(r)}_{-,p,\Delta,\ell}+P_{-}^{\mf b\mf c}\mathfrak{F}^{(r)}_{+,p,\Delta,\ell}; r\in \mathbb{Z}_{\ge 0}  \big\}\,,
	\ee
	and similarly with $\widetilde{\mathfrak{F}}$.
	The quantities $\mathfrak{F}^{(r)}_{-,p,\Delta,\ell}$ were defined in \eqref{eq:Polycond2}. It should be noted that these simple functionals are a variation of those discussed in \cite{Ghosh:2021ruh}. As we use only bosonic Witten diagrams,  in the language of 1d functionals it is like using `bosonic functionals' for the $-$ case and `fermionic functionals' for the $+$ case. They are not dual to any interesting CFT e.g. GFF, although they are perfectly good functionals for global symmetry problems. 
	
	The idea of simple functionals can be extended to QFTs with global symmetries. Here the $\pm$ functional actions should be replaced with the locality constraints of the $\mathcal{M}_\pm(s_1,s_2)$. We leave this for future work.

	\section{Conclusion}\label{sec: conclusion}
	
	In this paper we show that a crossing antisymmetric correlator can be expanded in a manifestly crossing antisymmetric basis. Similar to Polyakov's original proposal for crossing symmetric correlators, this basis is  mapped to AdS Witten diagrams.  We explore its relation to analytic functionals, and set up a dispersion relation that proves it.
	
	We call the building blocks of the basis the $+$ type Polyakov blocks. In 1d the simplest of these, the fermionic $+$ type Polyakov blocks, are constructed from a manifestly crossing antisymmetric sum of AdS$_2$ spin 1 exchange  bosonic  Witten diagrams. Their conformal block decomposition coefficients compute  1d fermionic $+$ type functional actions. The $+$ type functionals allow a  more relaxed Regge behavior compared to the $-$ type. This is used to obtain the bosonic Polyakov block by suitably combining spin 1 fermionic  Witten diagrams and a manifestly anticrossing function which we call a `contact diagram'. Once again under OPE decomposition  this exactly reproduces the bosonic $+$ functional action. 
	
	In general dimensions the $+$ type Polyakov blocks are built from odd spin Witten exchange diagrams and crossing antisymmetric contact diagrams. We formulate this in Mellin space. To obtain the exact structure of the blocks we introduce a crossing antisymmetric dispersion relation for a general crossing antisymmetric Mellin amplitude. This utilises a change of variables that results in nonlocal singularities. Imposing the locality constraints i.e. setting these  nonlocal terms to zero each odd spin sector of the dispersion relation reproduces a sum of odd spin exchange  Witten diagrams and finite number of contact diagrams. Finally we give a set of sum rules corresponding to the `Polyakov conditions' i.e. the residue at spurious double trace poles in Mellin variables should be zero. Using the dispersion relation the sum rules can be implemented for any spin. We have numerical checked with a simple example how locality conditions work, and how the Polyakov conditions are satisfied for the antisymmetric sector of a 2d WZW model, elaborately discussed in the Appendix.

	Our results imply a number of simplifications for bootstrapping 4-point functions where crossing antisymmetry becomes important. The most significant ones  being product functionals in even $d$ and various simple functionals for global symmetry problems in general $d$. There are many other interesting directions that one can pursue:
	
	\begin{enumerate}
	
		\item \textit{Constraints on QFTs}: A version of crossing symmetric dispersion relation has been used in	QFT context  \cite{Sinha:2020win}  to derive  constraints on effective field theories (equivalent to \cite{Caron-Huot:2020cmc, Tolley:2020gtv}) and scattering cross sections. They key element in that analysis was the locality constraints. These are  similar to our conditions \eqref{eq:localcondn} which are independent and should lead to new constraints on scattering amplitudes when they have a crossing antisymmetric component  e.g. when global symmetries are present.
		
		\item \textit{Geometric Function Theory}: An interesting new direction in constraining scattering amplitudes using ideas of Geometric Function theory (GFT) has been initiated in \cite{Haldar:2021rri,Raman:2021pkf}. The crossing symmetric dispersion relation allows a formulation in terms of `typically real functions' where GFT constraints  (e.g. Bieberbach-Rogonski inequalities) can be applied. In \cite{AhmadullahTA} this approach was extended to $O(N)$ theories building on the work of  \cite{Mahoux:1974ej}. It will be interesting to see how the crossing antisymmetric dispersion relation can be tied into the GFT framework and if the associated constraints/sum rules are connected to \cite{Mahoux:1974ej,AhmadullahTA} or independent ones. 
		
		\item \textit{Correlator bounds/Master functionals}: An interesting problem of obtaining bounds on CFT correlators was addressed in  \cite{Paulos:2020zxx, Ghosh:2021ruh}. To prove e.g. the minimization of 1d correlators one can construct a `master functional' which packages analytic functionals in a certain way. Interestingly the master functional action is equivalent to a Polyakov block.  It would be interesting to see how a crossing (anti)symmetric Polyakov block in higher dimensions is encoded in bounds (e.g. the numerical bounds obtained in \cite{Paulos:2021jxx} in 3d) on CFT correlators.

		\item \textit{Fermionic/Multiple correlator bootstrap}:  Crossing antisymmetry also appears in interesting problems like bootstrapping correlators of fermions in higher $d$ or a set of  correlators with unequal scalars \cite{Kos:2013tga,Iliesiu:2015qra,Poland:2018epd}. Our results are hence applicable to these cases (see e.g. \cite{Chowdhury:2021qfm}). However when fermions or unequal scalars are involved the conformal blocks are different, and also we may lose full crossing (anti)symmetry. So a modification of analytic functionals or Polyakov Bootstrap may be necessary.

			\item \textit{Nonperturbative Mellin amplitude/dispersive sum rules}: As shown in \cite{Gopakumar:2021dvg} the Polyakov conditions from crossing symmetric dispersion relation are equivalent to the analogous constraints proposed in \cite{Penedones:2019tng, Carmi:2020ekr} for nonperturbative Mellin amplitudes. These conditions can also be mapped to the ``dispersive sum rules" from position space dispersion relation or general $d$ analytic functionals \cite{Caron-Huot:2020adz}. It would be interesting to see if there is a similar picture for anticrossing correlators and if the associated sum rules are related to the crossing antisymmetric dispersion relation.

	\end{enumerate}

\section*{Acknowledgements}
The author acknowledges Subham Dutta Chowdhury, Kausik Ghosh, Rajesh Gopakumar, Miguel Paulos, Junchen Rong, Volker Schomerus, Aninda Sinha and Ahmadullah Zahed for many useful discussions. The author is thankful to Rajesh Gopakumar, Volker Schomerus and Aninda Sinha for their comments on the draft and various helpful suggestions. This work is supported by the German Research
Foundation DFG under Germany’s Excellence Strategy -- EXC 2121 ``Quantum Universe" -- 390833306.

	\appendix
	
	\section{Details of Witten diagrams}\label{app:Witten}

	In this section we give the details of the Witten diagrams used in the main text. The exchange Witten diagram $W_{\D,\ell}^{(s)}(u,v)$ with scalar external legs in general dimension i.e. AdS$_{d+1}$ is conveniently expressed in Mellin space as follows \cite{Gopakumar:2018xqi}
	\be\label{Wittengend}
	W_{\D,\ell}^{(s)}(u,v)=\int_{-i\infty}^{i\infty}[ds][dt]u^sv^t\Gamma^2(\Delta_\phi-s)\Gamma^2(-t)\Gamma^2(s+t)M^{(s)}_{\Delta,\ell}(s,t)\,,
	\ee
	 where $s,t$ are related to $s_i$ variables of section \ref{sec:gendPolyblock} by $s_1=s-\frac{2 \Delta _{\phi }}{3}$ and $s_2=\frac{\Delta _{\phi }}{3}+t$, and 
	 \begin{align}
	 M_{\Delta,\ell}(s,t)&=\widehat{P}_{\Delta-h,\ell}^{(s)}(s,t)\frac{\Gamma^2(\frac{\Delta+\ell}{2}+\Delta_\phi-h)}{(\frac{\Delta-\ell}{2}-s)\Gamma(\Delta-h+1)}\nonumber\\
	 &\times {}_3F_2\Big[\frac{\Delta-\ell}{2}-s,1+\frac{\Delta-\ell}{2}-\Delta_\phi,1+\frac{\Delta-\ell}{2}-\Delta_\phi;1+\frac{\Delta-\ell}{2}-s,\Delta-h+1;1\Big]\,.
	 \end{align}
	Here $h=d/2$ and we define $\widehat{P}_{\Delta-h,\ell}^{(s)}(s,t)$ below. The Mellin amplitude $M_{\Delta,\ell}(s,t)$ allows the decomposition
	\be\label{eq:mero}
	M_{\Delta,\ell}(s,t)=\Gamma^2(\frac{\D+\ell}2+\D_\phi-h)\widehat{P}_{\Delta-h,\ell}^{(s)}(s,t)\sum_{r=0}^{\infty}\frac{(1+\frac{\Delta-\ell}{2}-\Delta_\phi)_r^2(\frac{\Delta-\ell}{2}-s)}{r!\Gamma(\Delta-h+1+r)}\frac{1}{\frac{\Delta-\ell}{2}-s+r}\,.
	\ee
	The above form was used in e.g. \eqref{eq:dispersionmatch} to match the dispersion relation with the Witten diagrams. The coefficient $c_{\Delta,\ell}^{(k)}$ in \eqref{eq:branch} is given by
	\be
	c_{\Delta,\ell}^{(k)}=a_{\Delta,\ell} \, \mathcal{N}_{\Delta,\ell}\mathcal{R}_{\Delta,\ell}^{(k)}
	\ee
	where $a_{\Delta,\ell}$ is the squared OPE coefficient and
	\be
	\mathcal{N}_{\Delta,\ell}=\frac{(-2)^{\ell } (\Delta +\ell -1) \Gamma (\Delta-h +1) \Gamma (\ell +\Delta -1)^2}{\Gamma (\Delta -1) \Gamma^4 \left(\frac{\ell +\Delta }{2}\right) \Gamma^2 \left(\Delta_\phi-\lambda_2 \right) \Gamma^2 \left(\Delta_\phi+\lambda_1-h\right)}\,, \ \mathcal{R}_{\Delta,\ell}^{(k)}=\frac{\Gamma^2(\lambda_1+\Delta_\phi-h)(1+\lambda_2-\Delta_\phi)_k^2}{k!\Gamma(\Delta-h+1+k)}\,.
	\ee
Here $\lambda_1=(\Delta+\ell)/2$ and $\lambda_2=(\Delta-\ell)/2$\,.	

The $s$-channel Mack Polynomial $\widehat{P}_{\Delta-h,\ell}^{(s)}(s,t)$ is given by
\be
\widehat{P}_{\Delta-h,\ell}^{(s)}(s,t)=\sum_{m+n\le \ell}\mu^{(\ell)}_{m,n}\big(\frac{\Delta-\ell}{2}-s\big)_m(-t)_n
\ee	
where
\begin{align}
\mu^{(\ell)}_{m,n}&=\frac{2^{-\ell } \ell ! (-1)^{m+n}}{m! n! (\ell-m-n )!} \left(\lambda _1-m\right)_m \left(\lambda _2+n\right)_{\ell -n} (h+\ell -1)_{-m} (\ell +\Delta -1)_{n-\ell }  \left(\lambda _2+m+n\right)_{\ell-m-n }\nonumber \\
& \times \, _4F_3\left(-m,\lambda _2-h+1,\lambda _2-h+1,\Delta +n-1;\lambda _1-m,\lambda _2+n,\lambda_2 -2 h+2;1\right)\,.
\end{align}

To define the discontinuity and dispersion relation in section \ref{sec:gendPolyblock} we have also used the following shifted polynomials:
\be
P_{\Delta,\ell}(s_1,s_2)=\widehat{P}_{\Delta-h,\ell}^{(s)}(s_1+\frac{2\Delta_\phi}{3},s_2-\frac{\Delta_\phi}{3})\,.
\ee

To obtain expressions for 1d CFTs one should replace $u=z\zb$ and $v=(1-z)(1-\zb)$ in \eqref{Wittengend}\,.To obtain the deomposition into double trace blocks (see \eqref{eq:decomp} and \eqref{eq:decomp2}) of these exchange Witten diagrams one has to compute the residues of poles at $s=\Delta_\phi+n$\,.  In \cite{Gopakumar:2018xqi} it was shown that a convenient way of doing this is via the following expansion ($i=s,t,u$):
\be\label{eq:Mellindeco}
M^{(i)}_{\Delta,\ell}(s,t)=\sum_{\ell'}q^{(i)}_{\Delta,\ell'|\ell}(s) Q_{\ell',0}^{(2s+\ell')}(t)\,.
\ee
Here $Q_{\ell',0}^{(2s+\ell')}(t)$ is a continuous Hahn polynomial defined by
\be
Q_{\ell',0}^{(2s+\ell')}(t)=\frac{2^\ell((s)_\ell)^2}{(2s+\ell-1)_\ell}{}_3F_2[-\ell,2s+\ell-1,s+t;s,s;1]\,.
\ee
The coefficients $q^{(s)}_{\Delta,\ell'|\ell}(s)$ are given by rational functions, and computed using orthonormality of the Hahn polynomials. In the crossed channels we have  $q^{(t)}_{\Delta,\ell'|\ell}(s)$ that is given by a ${}_7F_6$ hypergoemetric function. See equations (3.6) and (3.28) of \cite{Gopakumar:2018xqi}. See also \cite{Ghosh:2021ruh} Appendix D on how to use recursively compute the coefficients of decomposition in double trace blocks using these relations. The decomposition \eqref{eq:Mellindeco} holds for contact diagrams i.e. $M_{\mathcal{C}}(s_1,s_2)$ (from \eqref{eq:Mellincont}) as well and the corresponding coefficients $q_{\ell'}$ are easy to compute rational functions.

\section{Derivation of the dispersion relation}\label{App:dispersion}

In this appendix we show the derivation of the crossing antisymmetric dispersion relation while also reviewing the crossing symmetric case following \cite{Sinha:2020win} (supplementary material). 

\subsection{General case - review}

\begin{figure}[t]
	\begin{center}
		\includegraphics[width=0.8\textwidth]{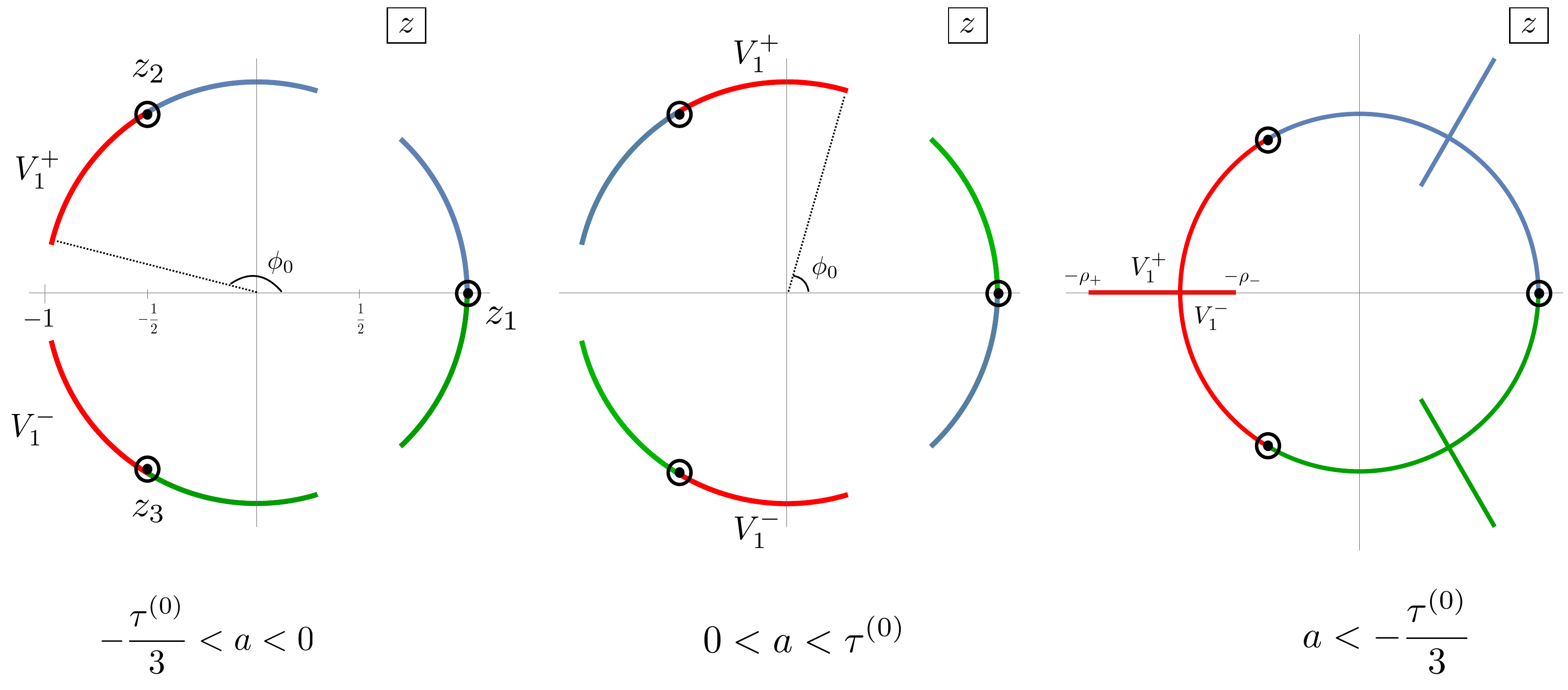}
	\end{center}
	\caption{
		A schematic picture of the physical cuts in $z$-plane. The colours indicate $V_1$ (red), $V_2$ (green) and $V_3$ (blue). 
		\label{zplane}}
\end{figure}

We first briefly review the derivation of a general dispersion relation without assuming crossing symmetry or antisymmetry. 
Let us recall the transformation of Mellin variables ($i=1,2,3$):
\begin{align}
s_i&=a-\frac{a(z-z_i)^3}{z^3-1},\label{stoza} \\  
a&=\frac{s_1s_2s_3}{s_1s_2+s_2s_3+s_1s_3}\label{stoa}\,,
\end{align}
with $z_i$ being cube roots of unity. The Mellin amplitude has a sequence of poles in $s_i\ge \tau^{(0)}$ that correspond to the physical operator content. \footnote{This is analogous to taking the branch cut $s_i\ge \frac{8m^2}{3}$ in a QFT scattering problem   where $s_i$ denotes Mandelstam variables and $m$ the mass of external particles.}
In terms of the new independent variables $z$ and $a$, these `physical cuts' correspond in the $z$ plane to the region $V_1(a)\cup V_2(a)\cup V_3(a)$ , as shown in Fig. \ref{zplane}, where
\begin{align}
V_1(a)&=\begin{cases}
&\{z;|z|=1,\frac 23 \pi \le |\text{arg }z|\le \phi_0(a)\}  \ \ \ \ \text{if }-\frac{\tau^{(0)}}{3}<a<0\,,\\
&\{z;|z|=1,  \phi_0(a) \le |\text{arg }z|\le \frac 23 \pi \}  \ \ \ \ \text{if } 0 <a<\tau^{(0)}\,,\\
&\{z;|z|=1,\frac 23 \pi \le |\text{arg }z|\le \pi  \}\cup \{z; \rho_-(a)\le |z|\le \rho_+(a),\text{arg }z=\pi \}  \ \  \text{if }a<-\frac{\tau^{(0)}}{3}
\end{cases}\nonumber\\
V_2(a)&=\exp\Big[\frac{2\pi i}{3}\Big]\, V_1(a)\\ 
V_3(a)&=\exp\Big[\frac{4\pi i}{3}\Big]\, V_1(a)\,,\nonumber
\end{align}
and we have defined 
\begin{align}
\phi_0(a)&=\tan^{-1}\bigg[\frac{\big[(\tau^{(0)}-a)(a+\frac{\tau^{(0)}}{3})\big]^{\frac 12}}{a-\frac{\tau^{(0)}}{3}}\bigg]\nonumber\\
\rho_{\pm}(a)&=\frac{3}{2\tau^{(0)}}\Bigg[ \bigg(\frac{\tau^{(0)}}{3}-a\bigg)  \pm \bigg[\big(\tau^{(0)}-a\big)\Big(-a-\frac{\tau^{(0)}}{3}\Big)\bigg]^{\frac 12} \Bigg]\,.
\end{align}
Notice that the cuts are defined only at $|z|=1$ or $\text{arg }z=\pi \text{ mod }\frac {2\pi}3$. These conditions correspond to $\text{Im }s_i=0$. Let us call the Mellin ampltude $\mathcal{M}(s_1,s_2)=\overline{\mathcal{M}}(z,a)$ in $z,a$ variables. Then discontinuities of $\overline{\mathcal{M}}(z,a)$ across the cut $V(a)$ are defined as
\be
\overline{\mathcal A}(z,a)=\begin{cases}
	\lim\limits_{\e\to 0} \frac{1}{2i\e}\big(\overline{\mathcal M}((1+\e)z,a)-\overline{\mathcal M}((1-\e)z,a)\big)\ \ \text{if } |z|=1\,, \\ \\ 
	\lim\limits_{\e\to 0} \frac{1}{2i\e}\big(\overline{\mathcal M}(ze^{i\e},a)-\overline{\mathcal M}(ze^{-i\e},a)\big)\ \ \text{if }\text{arg }z=\pi \text{ mod }\frac {2\pi}3\,.
\end{cases}
\ee
In terms of the  $\mathcal{A}_i(s_1,s_2)$ which is the usual discontinuity of $\mathcal{M}(s_1,s_2)$ defined across the cut in $s_i$, we have
\be\label{eq:AztoAs}
\overline{\mathcal A}(z,a)=\begin{cases}
	\pm \mathcal{A}_i(s_1,s_2) \ \ \forall \ z\in V_i^\pm(a) \ \ \text{ if }a<0\,,
	\\ 
	\mp \mathcal{A}_i(s_1,s_2) \ \ \forall \ z\in V_i^\pm(a) \ \ \text{ if }a>0 \,.
\end{cases}
\ee
Here we defined $V_i(a)=V_i^+(a)\cup V_i^-(a)$ such that $V_i^\pm=\{z;z\in V_i, \, \pm \text{Im\,} (z/z_i) \text{ or } \pm(|z|-1)< 0\}$ (shown in Fig. \ref{zplane})\,. 

To proceed one assumes the following Regge behavior of the Mellin amplitude: ${\mathcal{M}}(s_1,s_2)\sim O(s_i^{2-\e})$ (with $\e> 0$) at large $s_i$ \cite{Martin:1969ina} which translates to $\overline{\mathcal{M}}(z,a)\sim O((z-z_i)^{-2+\e})$ as $z\to z_i$. This implies that upon Taylor expansion $\overline{\mathcal{M}}(z,a)$ will have the form:
\be\label{eq:Taylor}
\overline{\mathcal{M}}(z,a)=\sum_{n=0}^{\infty}f_n(a)\, z^n\,.
\ee
that should converge for $|z|<\rho_-$ or $|z|<1$. 

Let us also add here the relation $\overline{\mathcal{M}}(z,a)=\overline{\mathcal{M}}^*(1/z^*,a)$ which simply follows from $\mathcal{M}(s_1,s_2)=\mathcal{M}^*(s_1^*,s_2^*)$\,. Note also that as $z\to 0$ or $z\to \infty$ one has $s_i\to 0$. Therefore $\overline{\mathcal{M}}(0,a)=\overline{\mathcal{M}}(\infty,a)=\overline{\mathcal{M}}^*(0,a)=\mathcal{M}(s_i=0)=f_0$, implying $f_0$ must be real and independent of $a$\,.

Now one considers the following contour integrals 
\begin{align}
I_-&=\frac{1}{2\pi i} \oint_{\mathcal{I}} dz' \frac{z'^3-1}{z'^3(z'-z)} \overline{\mathcal{M}}(z',a)=\frac{z^3-1}{z^3}\overline{\mathcal{M}}(z,a)+\frac{f_0}{z^3} +\frac{f_1(a)}{z^2}+\frac{f_2(a)}{z}\,,\label{eq:I-}\\
I_+&=\frac{1}{2\pi i} \oint_{\mathcal{E}} dz' \frac{z'^3-1}{z'^3(z'-z)} \overline{\mathcal{M}}(z',a)=f_0\,.\label{eq:I+}
\end{align}
The kernels are chosen to keep the integral finite near $z\to z_i$ and have poles at desired places.
Here $\mathcal{I}$ denotes the interior of the `cut' $V(a)$ where one computes the residues at  $z'=z$ (the first term of $I_-$) and $z'=0$ (for the other three terms). Also $\mathcal{E}$ denotes the exterior of $V(a)$ where one considers only the pole at $z'=\infty$. Then as we push $\mathcal{I}, \mathcal{E}$ towards $V(a)$ performing $I_+-I_-$ gives a dispersion relation with the discontinuity $ \overline{\mathcal{A}}(z,a)$ across $V(a)$, which may be written as
\be\label{eq:disp1}
\overline{\mathcal{M}}(z,a)=f_0+f_1(a)\frac{z}{1-z^3}+f_2(a)\frac{z^2}{1-z^3}+ \frac{z^3}{1-z^3}\frac{1}{\pi}\int_{V(a)} dz' \frac{z'^3-1}{z'^3(z'-z)} \overline{\mathcal{A}}(z',a)\,.
\ee

\subsection{Imposing crossing (anti)symmetry}

So far we have not used any assumption on full crossing symmetry or antisymmetry. In the notations of section \ref{sec:gendPolyblock} a crossing symmetric amplitude is always an expansion in $ x^py^q$ and for an anticrossing amplitude it should be in $ w x^py^q$ ($p,q\in \mathbb{Z}_{\ge 0}$). This implies the following behavior under the map \eqref{stoza}
\be\label{eq:cross}
\overline{\mathcal{M}}_{-}(z,a)\propto \bigg[\frac{{z^3}}{(z^3-1)^2}\bigg]^n\,, \ \ \overline{\mathcal{M}}_{+}(z,a)\propto \frac{{z^3(1+z^3)}}{(z^3-1)^3}\bigg[\frac{{z^3}}{(z^3-1)^2}\bigg]^n\,, \ n\ge 0\,.
\ee
where $+$ ($-$) denotes crossing (anti)symmetric case. Notice that both expression are functions of $z^3$. Hence the discontinuity $\overline{\mathcal{A}}_{+}(z,a)$ should be such that the general dispersion relation \eqref{eq:disp1} under small $z$ expansion would  allow only the powers $z^{3m}, \, m\in \mathbb{Z}_{\ge 0}$\,.
Therefore we may replace it with
\be\label{eq:disp2}
\overline{\mathcal{M}}_{\pm}(z,a)=f_{\pm,0}+ \frac{z^3}{(1-z^3)\pi}\int_{V(a)} dz' \frac{z'^3-1}{z'(z'^3-z^3)} \overline{\mathcal{A}}_{\pm}(z',a)\,,
\ee
where $f_{+,0}=0$. The above equation has an expansion in $z^{3m}$ and is equal to  \eqref{eq:disp1} up to those powers.  Indeed it can be checked that in \eqref{eq:I-}  and \eqref{eq:I+} if we began with the kernel of \eqref{eq:disp2} we would end up with the same result.

Now let us map this formula back to the variables $s_i$. For this we use \eqref{stoza}  to solve $z'$ in terms of $s_1'$ which gives two solutions $z'(s_1')$ and $z'^*(s_1')$ related by complex conjugation. \footnote{Using any  $s'_i$ ($i=1,2,3$) leads to the same result using croossing (anti)symmetry.)} We also focus on the case $-\frac{\tau^{(0)}}{3}<a<0$ where we have $|z'(s_1')|=1$, which further implies $z'^*(s_1')=1/z(s_1')$. Under this transformation the integrand of \eqref{eq:disp2} becomes
\be\label{eq:integrand}
\frac{dz'(s_1')}{ds_1'} \, \frac{z'(s_1')^3-1}{z'(s_1')(z'(s_1')^3-z^3)} \, \overline{\mathcal{A}}_{\pm}(z',a)+ \frac{dz'^*(s_1')}{ds_1'} \, \frac{z'^*(s_1')^3-1}{z'^*(s_1')(z'^*(s_1')^3-z^3)} \,  \overline{\mathcal{A}}_{\pm}(z'^*,a)\,
\ee
which is to be integrated over $s_1'$ and on $s_1'\ge \tau^{(0)}$. 
We now also solve $s_2'$ in terms of $s_1'$ and $a$ using \eqref{stoa} (recall that $a$ is fixed) which gives two solutions $s_2^{\pm}(s_1',a)$. We choose to use $s_2^{+}$\,. \footnote{Choosing between $s_2^+$ and $s_2^-$ corresponds to the convention of writing the branch cut. E.g. in writing \eqref{eq:branch} for odd $\ell$ the correct choice is $s_2^+$.}

Now we point out that we must have (using \eqref{eq:AztoAs})
\be
\overline{\mathcal{A}}_\pm(z'(s_1'),a)= \overline{\mathcal{A}}_\pm(z'^*(s_1'),a)=\mp {\mathcal{A}}_\pm(s_1',s_2^+(s_1,a))\,.
\ee
Note the `$-$' sign in the second equality for the crossing antisymmetric case. The change/invariance under $z\to z^*= 1/z$ is simple to see using \eqref{eq:cross} (as the $z$ dependencies are also shared by $\overline{\mathcal A}$). Now putting in the factor $\frac{z^3}{1-z^3}$ in \eqref{eq:integrand} we get for the crossing symmetric case (writing $f_{-,0}=\a_0$)
\be
\mathcal{M}_-(s_1,s_2)=\a_0+\frac{1}{\pi}\int ds_1' \, \mathcal{A}_-(s_1',s_2^+(s_1',a))\, H_-(s_1',s_1,s_2,s_3)\,,
\ee
with 
\be
H_-(s_1',s_1,s_2,s_3)=\frac{s_1}{s_1'-s_1}+\frac{s_2}{s_1'-s_2}+\frac{s_3}{s_1'-s_3}\,.
\ee
This is the relation presented in \cite{Sinha:2020win, Gopakumar:2021dvg}. For the crossing antisymmetric case we get our result \eqref{eq:antidisp}.

\section{Checks with the dispersion relation}
\subsection{Derivation of Witten diagrams for $\ell=5$}\label{app:ell5}
The polynomial $P_{\Delta,\ell}(s_1,s_2)$ from \eqref{eq:branch} for $\ell=5$ may be written as
\be
P_{\Delta,\ell=5}(s_1,s_2)=\sum_{m+n\le 5} b_{m,n}s_1^m s_2^n\,.
\ee
This form is easily seen by evaluating the polynomial at simple values e.g. $\Delta_\phi=1, \D=2, d=4$\,. With the above one can write the $\ell=5$ component of the Mellin amplitude in \eqref{eq:Mellin} as
\be
\mathcal{M}_{\Delta,\ell=5,k}(s_1,s_2)=H(\tau_k,s_1,s_2,s_3)\frac{1}{\tau_k}\sum_{m+n\le 5} b_{m,n}(\tau_k)^m s_2^+(s,a)^n=w\sum_{p=0}^\infty\sum_{q=0}^\infty M_{p,q,k}x^p a^q\,.
\ee
In the last equality we have expanded in all powers of $x$ and $a$. 

Let us also define the following quantity
\be
\mathcal{M}^{(0)}_{\Delta,\ell=5,k}(s_1,s_2)=\frac{P_{\Delta,5}(s_1,s_2)}{s_1-\tau_k}-\frac{P_{\Delta,5}(s_2,s_1)}{s_2-\tau_k}-\frac{P_{\Delta,5}(s_3,s_2)}{s_3-\tau_k}\,,
\ee
with $s_3=-s_1-s_2$. This is equivalent to the meromorphic part \eqref{eq:mero} of exchange Witten diagrams in all three channels. It is easy to see that the above can be written as 	
\be
\mathcal{M}^{(0)}_{\Delta,\ell=5,k}(s_1,s_2)=w\sum_{p=0}^2\sum_{q=0}^1 M^{(0)}_{p,q,k}x^{p+q} a^q\,.
\ee
We then find that the difference 
\be\label{eq:ell5}
\mathcal{M}_{\Delta,5,k}(s_1,s_2)-\mathcal{M}^{(0)}_{\Delta,5,k}(s_1,s_2)=w\Bigg[\Big\{ M_0 +\frac{x}{64\tau_k}+\frac{a x}{64\tau^2_k} \Big\}\, +\, \sum_{m>0}\widetilde{M}^{(1)}_m a^m+\sum_{n>1}\widetilde{M}^{(2)}_n a^n x\Bigg]\,.
\ee
Here $M_0=M_{0,0,k}-M^{(0)}_{0,0,k}$. The last two terms in \eqref{eq:ell5} are identified as nonlocal terms that should be set to zero. The terms in curly brackets are crossing antisymmetric contact diagrams that should be added to $\mathcal{M}^{(0)}_{\Delta,\ell=5,k}(s_1,s_2)$ to have the correct Polyakov block $\mathcal{P}_{\Delta,\ell}(u,v)$.

\subsection{Numerical checks}	\label{app:numchecks}
\subsubsection*{Locality conditions}

Here we discuss a numerical check of our discussions in section \ref{sec:gendPolyblock} with an explicit crossing antisymmetric solution. For this we consider the following artifical Mellin amplitude: 
\be
\mathcal{M}_0(s_1,s_2)=\frac{\left(s_1-s_2\right) \left(2 s_1+s_2\right) \left(s_1+2 s_2\right) \Gamma \left(-2 s_1-\frac{7}{8}\right) \Gamma \left(-2 s_2-\frac{7}{8}\right) \Gamma \left(2 s_1+2 s_2-\frac{7}{8}\right)}{\Gamma^2 \big(\frac{\Delta _{\phi }}{3}-s_1\big) \Gamma^2 \big(\frac{\Delta _{\phi }}{3}-s_2\big) \Gamma^2 \big(s_1+s_2+\frac{\Delta _{\phi }}{3}\big)}\,,
\ee
for $\D_\phi=\frac{21}{16}$ and $d=2$. It can be checked that at large $s_1$ we have $\mathcal{M}_0\sim s_1^{2/3}$ ($s_2$ fixed). The poles at $s_1=-\frac{7}{16}+m$, $m\in \mathbb{Z}_{\ge 0}$ gives operators at positions $\D=\frac 78 +n+\ell$ with $\ell$ odd.

We will verify the locality conditions \eqref{eq:localcondn} which can be done by assuming an expansion as follows:
\be\label{eq:aexp}
\mathcal{M}_0'(s_1,s_2)=\sum_{p=0}^{p_{\text{max}}}\sum_{q=0}^{q_{\text{max}}}\Bigg[\frac{s_2-s_3}{(-\frac 7{16}+\frac{p}{2})-s_1}+\frac{s_3-s_1}{(-\frac 7{16}+\frac{p}{2})-s_2}+\frac{s_1-s_2}{(-\frac 7{16}+\frac{p}{2})-s_3}\Bigg]a^q c_{p,q}\,,
\ee
with $s_3=-s_1-s_2$\,. The term in parentheses is an expansion in $w x^iy^j$. However the powers in $a$ introduce nonlocal terms. We can focus on the $p$-th pole of $s_1$ in $\mathcal{M}_0-\mathcal{M}'_0$ and expand around $s_2=0$\,. Solving them order by order one can obtain $c_{p,q}$\,. 

In Table \ref{numtab0} below we compare $\mathcal{M}_0(s_1,s_2)$ with $\mathcal{M}_0'(s_1,s_2)$ for $p_{\text{max}}=q_{\text{max}}=2$ and $p_{\text{max}}=q_{\text{max}}=8$ :
	\begin{table}[h]
	
	\begin{center}
		\begin{tabular}{|c|c|c|c|c|}
		\hline
			\ \ 		$s_1$ \ \ 	& \ \  $s_2$  \ \ & \ \   $\mathcal{M}_0$ \ \ & \ \   $\mathcal{M}'_0 \,  [p_{\max}= q_{\max}=2]$ \ \ & \ \  $\mathcal{M}'_0 \, [ p_{\max}=q_{\max}=8]$  \ \  \\ \hline
		  $\frac{i}{2}$ & $\frac{1}{14}$ & $-0.00899 - 0.18137 i$   & $-0.01089 - 0.19530 i$ & $-0.00915 - 0.18246 i$ \\ \hline
			$\frac{1}{10}$ & $\frac{1}{20}$ & $-0.056362$  & $-0.056168$ & $-0.056351$ \\ \hline
				$ \frac{1}{19} + \frac{i}{30}$ & $\frac{1}{10}(1+i)$ & $-0.00536 + 0.01651 i$ & $-0.00505 + 0.01645 i$ & $-0.00532 + 0.01650 i$ \\ \hline
		\end{tabular}
		\caption{Comparing $\mathcal{M}_0$ and $\mathcal{M}'_0$ for different choices of $s_1,s_2$ while increasing $p_{\max}$ and $q_{\max}$. The $s_1,s_2$ values were chosen to have a fast convergence. \label{numtab0}}
	\end{center}
\end{table}

This shows that it makes sense to work with variable $a$. Indeed the nonlocal singularities in \eqref{eq:aexp} must go to 0 which confirms the locality conditions. 

In order to check the precise disprsion formula \eqref{eq:antidisp} one needs to extract the OPE coefficients $a_{\D,\ell}$ from $\mathcal{M}_0$ using the Mellin integral \eqref{Mellinrep1}. For  a good match however one has to include operators with $n,\ell\sim 100$ (see \cite{Gopakumar:2021dvg}). We do not carry out this check. Instead we demonstrate the validity of Polyakov conditions \eqref{eq:Polycond} on a more physically relevant correlator. 

\subsubsection*{Polyakov conditions in WZW models}
We will consider $SU(2)_k$ WZW models in 2d. These theories have the global symmetry $SO(4)\cong \frac{SU(2)_L \times SU(2)_R}{Z_2}$. The 4-point function of scalars in a non-linear sigma model with a level-$k$ WZW term is known \cite{KNIZHNIK198483}. In \cite{He:2020azu}   the $O(4)$ sector of the correlator was studied for $k=1$. The same can be similarly obtained for any $k$. 

We will work with $k=2$ because it is simple and also our equations converge faster. The corresponding $O(4)$ sector is given by
\begin{align}
\frac{\mathcal{G}_{ijkl}(z,\zb)}{(z \zb)^{-\D_\phi}}=&P(\xi_\pm,\bar{\xi}_\pm)\Bigg[\bigg[\frac{\left(\bar{\xi }_-+\bar{\xi }_+\right){}^2}{\bar{\xi }_- \bar{\xi }_+}-\frac{\left(\xi _+-\xi _-\right) \left(\bar{\xi }_+-\bar{\xi }_-\right) \left(\bar{\xi }_- \bar{\xi }_++\xi _- \xi _+\right)}{\xi _- \xi _+ \sqrt{\xi _- \xi _++1} \bar{\xi }_- \bar{\xi }_+ \sqrt{\bar{\xi }_- \bar{\xi }_++1}}+\frac{\xi _+}{\xi _-}+\frac{\xi _-}{\xi _+}-2\bigg]\d_{ij}\d_{kl}\nonumber\\
&+ \bigg[\frac{\left(\xi _+-\xi _-\right) \left(\bar{\xi }_+-\bar{\xi }_-\right) \left(\frac{1}{\bar{\xi }_- \bar{\xi }_+}+\frac{1}{\xi _- \xi _+}+2\right)}{\sqrt{\xi _- \xi _++1} \sqrt{\bar{\xi }_- \bar{\xi }_++1}}-\frac{\bar{\xi }_+}{\bar{\xi }_-}-\frac{\bar{\xi }_-}{\bar{\xi }_+}-\frac{\xi _-}{\xi _+}-\frac{\xi _+}{\xi _-}+4\bigg]\d_{ik}\d_{jl}\nonumber\\
&+\bigg[\frac{\left(\xi _-+\xi _+\right) \left(\bar{\xi }_--\bar{\xi }_+\right)^2+\left(\xi _-+\xi _+\right) \left(\xi _--\xi _+\right)^2+\frac{2 \sqrt{z} \left(\bar{\xi }_+-\bar{\xi }_-\right) \left(\bar{\xi }_- \bar{\xi }_++\xi _- \xi _++2\right)}{\sqrt{\xi _- \xi _++1} \sqrt{\bar{\xi }_- \bar{\xi }_++1}}}{\xi _- \xi _+ \left(\xi _-+\xi _+\right) \bar{\xi }_- \bar{\xi }_+}\bigg]\d_{il}\d_{kj}\Bigg]\,,
\end{align}
where $\D_\phi=\frac 38$ and
\be
P(\xi_\pm,\bar{\xi}_\pm)=\frac{1}{16} \left(\xi _-+\xi _+\right) \left(\bar{\xi }_-+\bar{\xi }_+\right) ({\xi _- \xi _+ \bar{\xi }_- \bar{\xi }_+})^{\frac 14} \,, \ \xi_\pm=(1\pm \sqrt z)^{\frac 12}\,, \ \bar{\xi}_\pm=(1\pm \sqrt{\zb})^{\frac 12}\,.
\ee
As discussed around \eqref{eq:+typesol} we can use the parity odd eigenvector of the $O(N)$ crossing matrix (see \cite{Ghosh:2021ruh}) to obtain a crossing antisymmetric solution as follows:
\begin{align}
\frac{\mathcal{G}_+(z,\zb)}{(z \zb)^{-\D_\phi}}=\frac{\left(\xi _-+\xi _+\right) \left(\bar{\xi }_-+\bar{\xi }_+\right)}{16 \left(\xi _- \xi _+ \bar{\xi }_- \bar{\xi }_+\right){}^{3/4}} &\bigg[ 2 \xi _- \xi _+ \left(\bar{\xi }_-^2+\bar{\xi }_+^2\right)-\left(\xi _-^2+\xi _+^2\right) \left(\bar{\xi }_--\bar{\xi }_+\right)^2 \nonumber\\&-\frac{2 \left(\xi _--\xi _+\right) \left(\bar{\xi }_- \bar{\xi }_++\xi _- \xi _++1\right) \left(\bar{\xi }_--\bar{\xi }_+\right)}{{\left(\xi _- \xi _++1\right)^{1/2} \left(\bar{\xi }_- \bar{\xi }_++1\right)}^{1/2}}\bigg]\,.
\end{align}
This has a conformal block decomposition with dimensions $\D=n+\ell$ and spin $\ell$ for $n,\ell\in \mathbb{Z}_{\ge 0}$\,. 

\begin{figure}[t]
	\begin{center}
		\includegraphics[width=1\textwidth]{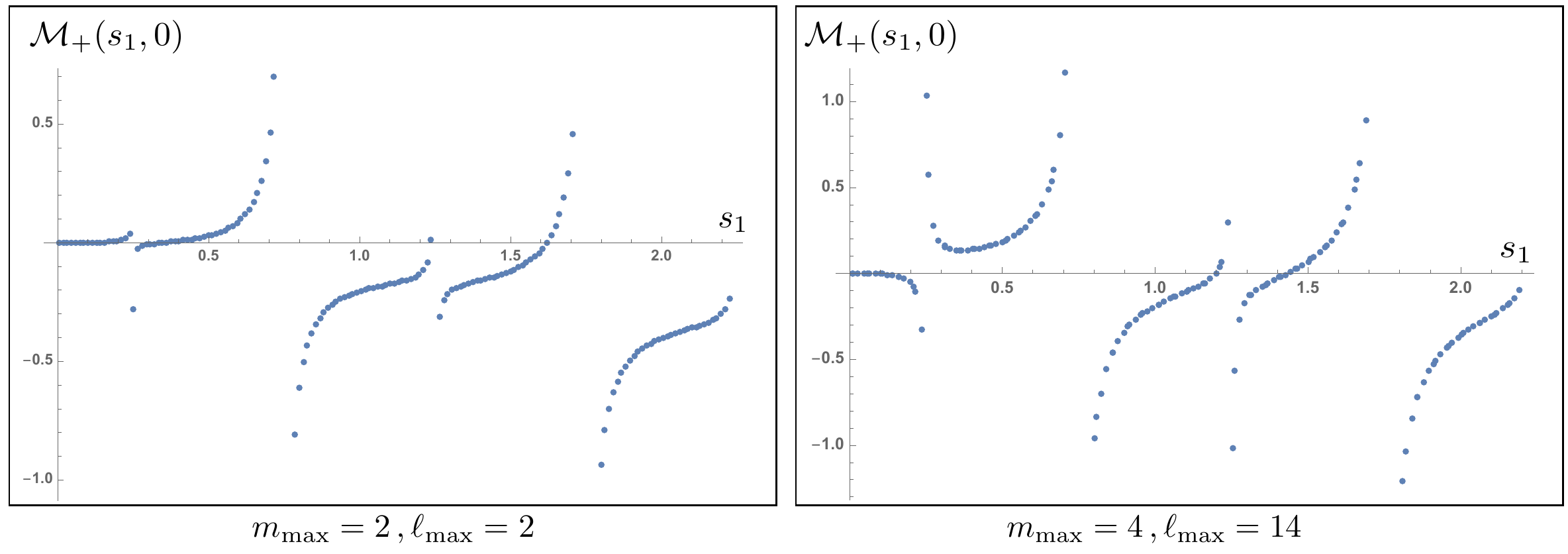}
	\end{center}
	\caption{
		The plot of Mellin amplitude $\mathcal{M}_+(s_1,s_2=0)$ for $m_{\max}=\ell_{\max}=2$ (left) and $m_{\max}=4\,, \ell_{\max}=14$ (right). The double zero structure at $s_1=\frac 18+p$ become prominent on adding more operators. 
		\label{Mellinplot}}
\end{figure}

We will test the conditions  $\mathfrak{F}_{+,p}(0)\equiv \mathcal{M}_+(s_1=\frac{1}{8}+p,s_2=0)=0$ from \eqref{eq:Polycond}\,. For this we use the above $\D,\ell$ to write  an approximate anticrossing Mellin amplitude as follows
\be
\mathcal{M}_+(s_1,s_2)=\sum_{\ell=0}^{\ell_{\text{max}}}\sum_{m=0}^{m_{\text{max}}}H_+\Big(\text{\small$\frac m2-\frac 14$},s_1,s_2,-s_1-s_2\Big)\sum_{n=0}^{2m}\frac{c^{(m-\frac{n}{2})}_{\D,\ell}}{(\text{\small$\frac m2-\frac 14$})}Q_{\ell,m-\frac{n}{2}}^{(\D)}(a)\,.
\ee
Here $\ell_{\text{max}}$ is a cutoff in spin and $m_{\text{max}}$ is the highest pole in $s_i$\,. In Table \ref{numtab} we show  $\mathfrak{F}_{+,p}(0)$ for diferent values of $\ell_{\text{max}}$, $m_{\text{max}}$\,. 
	\begin{table}[h]
	
	\begin{center}
		\begin{tabular}{|c|c|c|c|c|}
			\hline
		\ \ 	$\ell_{\max}$ \ \ 	& \ \ $m_{\max}$  \ \ & \ \ $\mathfrak{F}_{+,0}(0)$ \ \ & \ \  $\mathfrak{F}_{+,1}(0)$ \ \ & \ \  $\mathfrak{F}_{+,2}(0)$  \ \  \\ \hline
			4 & 2 & -0.00024 & -0.173489 & -0.37102\\ \hline
			12 & 2 & -0.00622 & -0.08865 & -0.25913\\ \hline
			12 & 12 & -0.00621 & -0.08770 & -0.21302 \\  \hline
		\end{tabular}
		\caption{Values of $\mathfrak{F}_{+,p}$ with $p=0,1,2$ for different choices of upper limits $\ell_{\max}, m_{\max}$.\label{numtab}}
	\end{center}
\end{table}

As we add operators $\mathfrak{F}_{+,p}$ becomes smaller except $\mathfrak{F}_{+,0}$ which grows with $\ell_{\text{max}}$. The latter is not very surprising as we are evaluating $\mathcal{M}_+(s_1,s_2)$ for low values of $s_1,s_2$, which always has a trivial zero for $s_1=s_2=0$ (from the antisymmetric term $w$). On adding operators the different zeroes must decouple.

In Figure \ref{Mellinplot} we have plotted $\mathcal{M}_+(s_1,s_2=0)$ for $0\le s_1\le 2.2$\,. The nontrivial double zeroes are expected at $s_1=0.125+p$ and poles at $\frac{m}{2}-0.25$\,. The double zero locations get more accurate as we add more operators. As expected the convergence gets poorer for larger values of $s_1$. We carried out the above analysis also for level $k=3$, which yields similar results.

	\section{Product functionals in $d=4$ and $d=6$}\label{app:productfunc}
	
	In this appendix we review how to use the product functionals in $d=4$ and $d=6$. We then use the simple functionals discussed in section \ref{sec:simple} for 2d to obtain analytic bounds on OPE coefficients.
	
	\subsection{$d=4$}
	In $d=4$ we have the conformal blocks:
	\begin{align}
	G^{d=4}_{\D,\ell}(z,\bar z)&=\frac {1}{1+\ell} \frac{z \zb}{z-\zb}\big[G_{\frac{\rho}{2}}(z|\text{\tiny$\frac{\D_\phi}{2}$})G_{\frac{\tau-2}{2}}(\bar z|\text{\tiny$\frac{\D_\phi}{2}$})-(z\leftrightarrow \bar{z})\big]\,,\nonumber\\
	&=\frac {1}{1+\ell} \frac{1}{z-\zb}\big[G_{\frac{\rho}{2}}(z|\text{\tiny$\frac{\D_\phi-1}{2}$})G_{\frac{\tau-2}{2}}(\bar z|\text{\tiny$\frac{\D_\phi-1}{2}$})-(z\leftrightarrow \bar{z})\big]\,.
	\end{align}
	The crossing vector becomes
	\be
	F^{d=4}_{-,\D,\ell}(z,\bar z|\D_\phi)=\frac{\big[ F_{+,\frac{\rho}{2}}(z|\text{\tiny$\frac{\D_\phi-1}{2}$})F_{+,\frac{\tau-2}{2}}(\bar z|\text{\tiny$\frac{\D_\phi-1}{2}$})+ F_{-,\frac{\rho}{2}}(z|\text{\tiny$\frac{\D_\phi-1}{2}$})F_{-,\frac{\tau-2}{2}}(\bar z|\text{\tiny$\frac{\D_\phi-1}{2}$}) -(z\leftrightarrow \bar z)\big]}{2(1+\ell)(z-\zb)} \,.
	\ee
	We can combine $\mathcal{I}_z\mathcal{I}_{\zb}F_{\D,\ell}(z,\zb)$ and $\mathcal{I}_z\mathcal{I}_{\zb}F_{\D,\ell}(z,1-\zb)$ to get expressions involving only $F_+$ or only $F_-$. So
	the product functionals must involve either both $+$ type functionals or both $-$ type.
	We choose the functional to be 
	\be
	[\omega^{(1)}_{\pm}\otimes\omega^{(2)}_{\pm}]^{d=4}(\D,\ell|\D_\phi)=2\int_{1}^{\infty}\frac{dz d{\bar z}}{\pi^2}h^{(1)}_{\pm}(z)h^{(2)}_{\pm}(\bar z)(z-\zb)\big[\mathcal{I}_z\mathcal{I}_{\bar z}F_{-,\D,\ell}(z,\bar z)\pm\mathcal{I}_z\mathcal{I}_{\bar z}F_{-,\D,\ell}(z,1-\bar z)\big]\,.
	\ee
	The kernels are chosen from $\a_{\pm,n},\b_{\pm,n}$\,. However notice the $z-\zb$ factor in the inegral above. It means we have to pick a combination of functional that have a stronger Regge falloff and cancels this factor. Let us work with fermionic functionals $\a^F_{\pm,n},\b^F_{\pm,n}$ and pick
	\begin{align}\label{eq:subtract}
	\tilde \b^F_{-,n}&=\b^F_{-,n}-b_n \b^F_{-,0} \,, \hspace{1cm} \text{$b_n$ chosen such that} \hspace{0.5cm} f_{\tilde \b^F_{-,n}}=O(z^{-4})\,,\nonumber\\
	\tilde \a^F_{-,n}&=\a^F_{-,n}-a_n \b^F_{-,0} \,, \hspace{1cm} \text{$a_n$ chosen such that} \hspace{0.5cm} f_{\tilde \a^F_{-,n}}=O(z^{-4})\,.
	\end{align}
	Note that for bosonic functions (recall $\beta_{-,0}^B=0$) we should subtract $\a^B_{-,0}$ instead.
	 The product functionals are then as follows (denoting $\omega^{(1)}\omega^{(2)}\equiv [\omega^{(1)}\otimes\omega^{(2)}]^{d=4}$ and suppressing $F$ superscripts)
	\begin{align}
	&\b_{+,m}\b_{+,m}\hspace{1cm}m,n\ge 0\,, m\ne n\,, \hspace{1cm}\a_{+,m}\a_{+,m}\hspace{1cm}m,n\ge 0\,, m\ne n\,,\nonumber\\&\b_{+,m}\a_{+,m}\hspace{1cm}m,n\ge 0\,, \hspace{2.4cm}\tilde\b_{-,m}\tilde\b_{-,m}\hspace{1cm}m,n\ge 1\,, m\ne n\,,\nonumber\\&\tilde\a_{-,m}\tilde\b_{-,m}\hspace{1cm}m,n\ge 0\,, m\ne n\,, \hspace{1cm}\tilde\b_{-,m}\tilde \b_{-,m}\hspace{1cm}m\ge 1\,,n\ge 0\,.
	\end{align}
	The functional action is given by
	\be
	\omega^{(1)}_{\pm}\omega^{(2)}_{\pm}(\D,\ell|\D_\phi)=\frac{1}{\ell+1}\big[\omega_\pm^{(1)}(\rho|\D_\phi|4)\omega_\pm^{(2)}(\tau-2|\D_\phi|4)-\omega_\pm^{(2)}(\rho|\D_\phi|4)\omega_\pm^{(1)}(\tau-2|\D_\phi|4)\big]\,.
	\ee
	Here we defined
	$\omega_{\pm}(\D|\D_\phi|4)=\omega_{\pm}(\text{\tiny$\frac{\D}{2}$}|\text{\tiny$\frac{\D_\phi-1}{2}$})$\,. As in 2d the action is entirely determined from the $\pm$ type Polyakov blocks. 
	
	However one should be careful as subtractions \eqref{eq:subtract} are involved in the $-$ case. We saw  in section \ref{sec:bosonic} (for the $+$ case) that to obtain Polyakov blocks related to functionals with stricter large $z$ fall-off one has to add more contact diagrams. Therefore to obtain $\tilde{\beta}_{-,n}, \tilde{\alpha}_{-,n}$ one has to add a contact diagram to $\mathcal{P}_{-,\Delta}(z)$ such that one coefficient in its conformal block decomposition is zero  \cite{Mazac:2018ycv}.

	\subsection{$d=6$}
	The $d=6$ conformal blocks are given by
	\begin{align}
	G_{\D,\ell}^{d=6}(z,\zb)&=K\frac{z^2 \bar{z}^2}{(z-\zb)^2}\Big[G_{\D-2,\ell}^{d=4}(z,\zb)-G_{\D-2,\ell+2}^{d=4}(z,\zb)\nonumber\\
	&-\frac{(\Delta -4) (\tau -4)^2}{16 (\Delta -2) \left(\tau ^2-8 \tau +15\right)} G_{\D,\ell}^{d=4}(z,\zb)+\frac{(\Delta -4) \rho ^2}{16 (\Delta -2) \left(\rho ^2-1\right)}G_{\D,\ell+2}^{d=4}(z,\zb)\Big]\,.
	\end{align}
	where
	\be
	K=-\frac{6 \left(\rho ^2-1\right) \left(\tau ^2-8 \tau +15\right) }{(\rho -1) (\rho +1) (\tau -5) (\tau -3) (\ell +2) }\,.
	\ee
	The crossing vector is given by
	\begin{align}
	F_{\D,\ell}^{d=6}(z,\zb|\D_\phi)&=\frac{K}{ \left(z-\bar{z}\right)^2}\Big[F_{\D-2,\ell}^{d=4}(z,\zb|\D_\phi-2)-F_{\D-2,\ell+2}^{d=4}(z,\zb|\D_\phi-2)\nonumber\\
	&-\frac{(\Delta -4) (\tau -4)^2}{16 (\Delta -2) \left(\tau ^2-8 \tau +15\right)} F_{\D,\ell}^{d=4}(z,\zb|\D_\phi-2)+\frac{(\Delta -4) \rho ^2}{16 (\Delta -2) \left(\rho ^2-1\right)}F_{\D,\ell+2}^{d=4}(z,\zb|\D_\phi-2)\Big]\,.
	\end{align}
	Similar to the $d=4$ case we choose the functional kernels in the following way:
	\be\label{d6int}
	[\omega^{(1)}_{\pm}\otimes\omega^{(2)}_{\pm}]^{d=6}(\D,\ell|\D_\phi)=2\int_{++}\frac{dz d{\bar z}}{\pi^2}h^{(1)}_{\pm}(z)h^{(2)}_{\pm}(\bar z)(z-\zb)^3\big[\mathcal{I}_z\mathcal{I}_{\bar z}F_{\D,\ell}(z,\bar z)\pm\mathcal{I}_z\mathcal{I}_{\bar z}F_{\D,\ell}(z,1-\bar z)\big]\,.
	\ee
	The functional action is given by ($\omega^{(1)}\omega^{(2)}\equiv \big[\omega^{(1)}\otimes\omega^{(2)}\big]$)
	\begin{align}
	&[\omega^{(1)}_{\pm}\omega^{(2)}_{\pm}]^{d=6}(\D,\ell|\D_\phi)=K\Big[[\omega^{(1)}_{\pm}\omega^{(2)}_{\pm}]^{d=4}(\D-2,\ell|\D_\phi)-[\omega^{(1)}_{\pm}\omega^{(2)}_{\pm}]^{d=4}(\D-2,\ell+2|\D_\phi)\nonumber\\
	&-\frac{(\Delta -4) (\tau -4)^2 [\omega^{(1)}_{\pm}\omega^{(2)}_{\pm}]^{d=4}(\D,\ell|\D_\phi) }{16 (\Delta -2) \left(\tau ^2-8 \tau +15\right)} +\frac{(\Delta -4) \rho ^2[\omega^{(1)}_{\pm}\omega^{(2)}_{\pm}]^{d=4}(\D,\ell+2|\D_\phi)}{16 (\Delta -2) \left(\rho ^2-1\right)} 
	\Big]
	\end{align}
	Since now we have a $(z-\zb)^3$ in the integral of \eqref{d6int} we have to subtract more functionals while choosing the kernels, so that the falloff is at least $O(z^{-5})$. To have a Regge bounded integrand we should subtract one functionals from the $+$ type ones and 2 from the $-$ type. Working with fermionic functionals we pick:
	\begin{align}
	\hat \b_{-,n}&=\b_{-,n}-b_n^{(1)} \b_{-,0} -b_n^{(2)} \a_{-,0}\,, \hspace{1cm} \text{$b_n^{(1,2)}$ chosen such that} \hspace{0.5cm} f_{\hat \b_{-,n}}=O(z^{-6})\,,\nonumber\\
	\hat \a_{-,n}&=\a_{-,n}-a_n^{(1)} \b_{-,0} -a_n^{(2)} \a_{-,0}\,, \hspace{1cm} \text{$a_n^{(1,2)}$ chosen such that} \hspace{0.5cm} f_{\hat \a_{-,n}}=O(z^{-6})\,,\nonumber\\
	\hat \b_{+,n}&=\b_{+,n}-c_n \b_{+,0} \,, \hspace{3cm} \text{$c_n$ chosen such that} \hspace{0.5cm} f_{\hat \b_{+,n}}=O(z^{-5})\,,\nonumber\\
	\hat \a_{+,n}&=\a_{+,n}-d_n \b_{+,0} \,, \hspace{3cm} \text{$d_n$ chosen such that} \hspace{0.5cm} f_{\hat \a_{+,n}}=O(z^{-5})\,.
	\end{align}
	To obtain their actions from Witten diagrams,  we should appropriately add two (crossing symmetric) contact diagrams to $\mathcal{P}_{-,\Delta}(z)$ and one (crossing antisymmetric) contact diagram to $\mathcal{P}_{+,\Delta}(z)$.
	
	The $d=6$ product functionals are hence chosen as follows
	\begin{align}
	&\hat \b_{+,m}\hat \b_{+,m}\hspace{1cm}m,n\ge 1\,, m\ne n\,, \ \hspace{1cm} \hat \a_{+,m} \hat \a_{+,m}\hspace{1cm}m,n\ge 0\,, m\ne n\,,\nonumber\\&\hat \b_{+,m}\hat \a_{+,m}\hspace{1cm}m\ge 1\,,n\ge 0\,, \ \hspace{1.5cm} \hat\b_{-,m}\hat\b_{-,m}\hspace{1cm}m,n\ge 2\,, m\ne n\,,\nonumber\\&\hat\a_{-,m}\hat\b_{-,m}\hspace{1cm}m,n\ge 0\,, m\ne n\,,\ \hspace{1cm} \hat\b_{-,m}\hat \b_{-,m}\hspace{1cm}m\ge 2\,,n\ge 0\,.
	\end{align}
	\subsection{Bounds with product functionals}\label{sec:bounds}
	
    In this subsection we use the simple functionals built with the 2d product functionals (see section \ref{sec:productfunc2d}) to extend the bounds on OPE density obtained in $d=2$ \cite{Paulos:2019gtx} to a CFT with  general global symmetry. 
	For the uncharged case the bound was obtained by choosing the following combination of 1d functionals:
	\be
	\tilde{\a}_{\pm,m}=\a_{\pm,m} + c_{\pm}\b_{\pm,m}, \ \ \text{$c_{\pm}$ chosen such that} \ \ f_{\a_{\pm,m}}\sim O(z^{\frac 12(9\pm 1)})\,.
	\ee
	These modified functionals have the following positivity and asymptotic properties respectively:
	\begin{align}
	&\tilde{\a}_{+,m}(\tau)\ge 0, \ \ \forall \ \tau \ge 0\,, \ \ \ \ \tilde{\a}_{-,m}(\tau)\ge 0, \ \ \forall \ \tau \ge \tau_0(\D_\phi)\,,\\
	&\tilde{\a}_{\pm,m}(0)\stackrel{m\to \infty}{=}\pm a_{h_m}^{\text{free}}\,,
	\end{align}
	where
	\be
	a_{h}^{\text{free}}=\frac{2\Gamma(\frac h2)^2}{\Gamma(h-1)\Gamma(\D_\phi)^2}\frac{\Gamma(\frac{h+2\D_\phi-2}{2})}{\Gamma(\frac{h+2\D_\phi+2}{2})}\,, \ \ \ \ h_m=2+2\D_\phi+4m\,.
	\ee
	For $h,h_m\gg 1$ with $h-h_m$ fixed we have
	\be
	\a_{\pm,m}(h)\stackrel{h,h_m\to \infty}{=}\left(\frac{a_{h_m}^{\text{free}}}{a_h^{\text{free}}}\right)\left(\frac{4}{\pi}\frac{\sin(\frac{\pi}{4}(h-h_m))}{h-h_m}\right)^2\,.
	\ee
	Now the action of the product functional $\tilde{\a}_{-,m}\tilde{\a}_{+,n}$ on the bootstrap equation implies the following bound:
	\be
	\sum_{\stackrel{|\D-\D^{\text{prod}}_{m,\ell}|\le 2}{\ell=2(m-n)}}(1+\d_{\ell,0})a_{\D,\ell}\tilde{\a}_{-,m}(\tau)\tilde{\a}_{+,n}(\rho)\le -\sum_{\stackrel{0\le \tau\le \tau_0}{\ell=0,2,\cdots}}a_{\D,\ell}\big[\tilde{\a}_{-,m}(\tau)\tilde{\a}_{+,n}(\rho)+\tilde{\a}_{+,n}(\tau)\tilde{\a}_{-,m}(\rho)\big]\,.
	\ee
	Here $\D_{m,\ell}^{\text{prod}}=2+2\D_{\phi}+4m+\ell$\,.
	Using the above properties of $\tilde{\a}_{\pm,m}$ we get the following upper bound on OPE density:
	\be\label{opebound}
	\lim\limits_{m,n\to \infty}\sum_{\stackrel{|\D-\D^{\text{prod}}_{m,\ell}|\le 2}{\ell=2(m-n)}}\left(\frac{a_{\D,\ell}}{a_{\D,\ell}^{\text{prod}}}\right)\left(\frac{4}{\pi}\frac{\sin(\frac{\pi}{4}(\D-\D_{m,\ell}))}{\D-\D_{m,\ell}}\right)^4\le 1\,,
	\ee
	where $a_{\D,\ell}^{\text{prod}}=\frac{2}{1+\d_{\ell,0}}a_{\rho}^{\text{free}}a_{\tau}^{\text{free}}$\,. 
	
	Recall from section \ref{sec:simple} that the global symmetry problem is formulated in terms of a crossing matrix $C^{\mathfrak{ab}}$. It has the eigenvalues $+1$ with $r_+$ degeneracy and $-1$ with $r_-$ degeneracy. 
	It was shown in \cite{Ghosh:2021ruh} that there can be  $r_+-r_-$ eigenvectors with $+1$ eigenvalue, that  have positive even parity components and zero parity odd components. These are found by taking linear combinations of them and   ($\textbf S$ : singlet):
	\be
	E_{+}^{\mf a}=\begin{cases}
		\frac{2+d_{\mf r}}{2}, & \mf a=\textbf{S} \\
		\sqrt{d_{\mf r}}, & \eta^{\mf a}=1 \\
		0, & \eta^{\mf a}=-1 
	\end{cases}\,, \ \ 
	\tilde{E}_{+}^{\mf a}=\begin{cases}
		\frac{d_{\mf r}}{2}, & \mf a=\textbf{S} \\
		\sqrt{d_{\mf r}}, & \eta^{\mf a}=1 \\
		0, & \eta^{\mf a}=-1 
	\end{cases}\,,
	\ee
	which are also $+1$ eigenvectors (and $\eta$ is defined by the tensor $T^{\mf a}_{ij,kl}=\eta^{\mf a}T^{\mf a}_{ji,kl}$, see \eqref{eq:Gijkl}). 
	In fact the other linearly independent $r_-$ combinations can also be chosen to have non-negative components. 
	
	Now we note that for any eigenvector $e_{\pm}$ we must have $\sum_{\mf b} e_{\pm}^{\mf b}P_{\mp}^{\mf{a}\mf{b}}=0$. 
	Choosing a non-negative eigenvectors as above (lets say $E^{\mf b}_{+,s}$), we can consider the combination of simple functionals $E_{+,s}^{\mf b}\omega^{\mf b|\mf c}$ whose action would be proportional to $\omega_-^{(1)}\omega_-^{(2)}(\D,\ell)$. Choosing  $\omega_-^{(1)}\omega_-^{(2)}=\tilde{\a}_{-,m}\tilde{\a}_{+,n}$ its action translates the OPE denisity upper bound \eqref{opebound}  to the following bound for global symmetry:
	\be
	\lim\limits_{m,n\to \infty}\sum_{\stackrel{|\D-\D^{\text{prod}}_{m,\ell}|\le 2}{|\ell-2(m-n)|\le 1}}\left(\frac{\sum_{\mf a}E^{\mf a}_{+,s}a^{\mf a}_{\D,\ell}}{a_{\D,\ell}^{\text{prod}}}\right)\left(\frac{4}{\pi}\frac{\sin(\frac{\pi}{4}(\D-\D_{m,\ell}))}{\D-\D_{m,\ell}}\right)^4\le \sum_{\mf b} E^{\mf b}_{+,s} a_0^{\mf b}\,.
	\ee
	Here $a_0^{\mf a}$ denotes the OPE of $\D=0,\ell=0$ operator in each irrep $\mf a$ (with $a^{\textbf S}_0=1$). Since all components $E_{+,s}^{\mf a}$ are positive this implies the following individual bound:
	\be
	\lim\limits_{m,n\to \infty}\sum_{\stackrel{|\D-\D^{\text{prod}}_{m,\ell}|\le 2}{|\ell-2(m-n)|\le 1}}\left(\frac{E^{\mf a}_{+,s}a^{\mf a}_{\D,\ell}}{a_{\D,\ell}^{\text{prod}}}\right)\left(\frac{4}{\pi}\frac{\sin(\frac{\pi}{4}(\D-\D_{m,\ell}))}{\D-\D_{m,\ell}}\right)^4\le \frac{1}{E_{+,s}^{\mf a}}\sum_{\mf b} E^{\mf b}_{+,s} a_0^{\mf b}\,.
	\ee


	\small
	\parskip=-10pt
	\bibliography{mybib}
	\bibliographystyle{jhep}

\end{document}